\documentstyle{article}
\begin{document}

\title{On the stochastic dynamics of disordered spin models}
\author{G. Semerjian$^1$, L. F. Cugliandolo$^{1,2}$, A. Montanari$^1$\\
$^1$ Laboratoire de Physique Th\'eorique de l'Ecole Normale Sup\'erieure\\
24 rue Lhomond, 75231 Paris Cedex 05, France\\
$^2$ Laboratoire de Physique Th\'eorique et Hautes Energies, Jussieu\\
4, Place Jussieu, Tour 16, 1er \'etage, 75252 Paris Cedex 05, France}

\maketitle
\begin{abstract}
In this article we discuss several aspects of the stochastic dynamics of spin models. The paper has two independent parts. Firstly, we explore a few properties of the 
multi-point correlations and responses of generic systems evolving in equilibrium with a thermal bath. We propose a fluctuation principle that allows us to derive fluctuation-dissipation relations for many-time correlations and linear responses.  We also speculate on how these features will be modified in systems evolving slowly out of equilibrium, as finite-dimensional or dilute spin-glasses. Secondly, we present a formalism that allows one to derive a series of approximated equations that determine the dynamics of disordered spin models on random (hyper) graphs.
\end{abstract}

\section{Introduction}

There are several motivations to revisit the 
stochastic dynamics of spin models with and without disorder.
Starting from a random initial condition, at low enough temperature, these models usually 
have a very slow evolution with several aspects in common
with the one of real glassy systems. For instance, during coarsening 
the global auto-correlation functions of ferromagnetic Ising models on finite dimensional lattices~\cite{MartinEvans} age in a rather similar way to the one 
observed in molecular dynamic simulations of Lennard-Jones
mixtures~\cite{KobBarrat}, a typical glass former. The asymptotic linear response~\cite{ABarrat}
to an external perturbation of these non-frustrated and non-disordered  models is, however, different from the one measured in 
glasses. The non-trivial 
slow response observed numerically~\cite{KobBarrat} and experimentally~\cite{Struick} is 
captured by modified spin models in which frustration and/or disorder are added~\cite{yo}. A standard example is the fully-connected 
disordered $p$ spin model in which all $p$-uplets of spins interact 
via random exchanges. The statics~\cite{Kithwo} and Langevin dynamics~\cite{Cuku1} of 
this mean-field model  have been solved analytically in great detail and there is growing consensus in that this and related models yield a mean-field description of the  structural glass
transition and glassy dynamics. 

More recently, the interest in studying similar disordered spin models on random graphs and random hyper-graphs (as opposed to the complete graph or hyper-graph) has grown. 
The reasons for this are multiple. 

From the glassy point of view, the dilute disorder models ``approach'' the finite dimensional actual problem one would like to understand while still being mean-field. Due to the  dilute nature of the interactions not all fluctuations are suppressed in the thermodynamic limit. In more 
technical terms, when the number of spins in the system diverges, the disordered average dynamics is not completely described by the two-time global auto-correlation and linear response but 
all kinds of many-time functions carry non-trivial information 
about the dynamic behavior of these systems.  Since this will also happen in finite dimensional problems one might expect that the knowledge on the behavior of  these higher-order correlations 
in the dilute problem be of help.

Dilute disordered spin models are highly non-trivial even in 
their paramagnetic high-temperature phase. As shown by 
Bray~\cite{Bray-Griffiths} these models have a Griffiths phase 
``generated'' by the fluctuations in the connectivities of the 
vertices of the random (hyper) graph. A complete solution to the 
dynamics of one such model might help understanding the nature and properties of Griffiths phases. In particular, the existence or
not of a Griffiths phase in quantum disordered finite-dimensional
systems  in 
contact with a quantum environment has been the subject of
intense debate recently~\cite{Antonio}. The analytic solution of a quantum model even if defined on a random graph might be 
of help in this discussion.

Dilute disordered spin models yield also a representation
of several problems of great interest in computer science~\cite{tcs}. 
For instance, the dilute $p$ spin disordered ferromagnetic 
model represents the so-called {\sc xor}-sat problem~\cite{xor} and 
variations on the $p$ spin disordered spin-glass model
describe the $k$-sat optimization problem~\cite{ksat,cavity}. 
These models 
are usually attacked with numerical algorithms that do not 
correspond to a physical dynamics like the Glauber or Langevin 
ones~\cite{dynamics-sat}. Having said this, it would be interesting and useful, also for these computational problems, to understand their evolution under 
physical dynamics. 

Finally, in several problems of physical interest, as the 
gelation process, dilute random matrices play an important 
role~\cite{Annette}. The methods and results here discussed 
will be relevant for these problems too. 

This paper presents two independent results for the 
Langevin dynamics of spin systems. In the first part
we derive several general relations between many-point
correlations and linear responses that any system in 
equilibrium with a thermal bath must respect. These equations are 
derived using a general fluctuation principle. We then 
discuss how these relations might be modified in a 
system evolving slowly out of equilibrium as we know 
the above mentioned models do when the temperature of the 
environment is sufficiently low. In the second part of the paper 
we introduce a functional method to attack the Langevin dynamics 
of spin models defined on random graphs. 
We discuss how one recovers the well-known
Schwinger-Dyson dynamic equations for  the disordered averaged two-time correlations and linear responses in the fully-connected 
limit. We next explain a set of iterations that allow us to deal with the 
disordered averaged dynamics of the dilute problems in an 
approximated way. This part extends results briefly presented in \cite{Secu}.

We wish to stress here that 
since we expect self-averageness when an infinite system evolves 
out of equilibrium, a model with a typical realization of 
disorder should behave in the  same way as the averaged 
result predicts. Still, we also know that the infinite size model 
will keep local fluctuations due to the sample-dependent
fluctuations in the site connectivities and the 
random exchanges that will be lost when performing the 
disorder average. 

The paper is organized as follows. In Section \ref{sectionmanypoint} we discuss
the properties of many-time correlations in and out of equilibrium.
We then argue on which are the modifications of these relations
(applied to global functions) that are  
expected in a system that slowly evolves out of equilibrium. 
In Section~\ref{dilute} we give a precise definition of the models 
we are interested in and we introduce the functional method. Finally, we present our conclusions and directions for future
work. 

\section{Many-point functions}
\label{sectionmanypoint}

A way to characterize the dynamics of a generic
model is to determine the evolution of the 
many-point correlators and linear responses defined as
\begin{eqnarray}
C(i_1,t_1,\dots,i_k,t_k) &=& \langle s_{i_1}(t_1) \dots s_{i_k}(t_k) \rangle
\;,
\\
R(i_1,t_1,\dots,i_{k-1},t_{k-1}; i_{k},t_k) &=& 
\left.
\frac{\delta \langle s_{i_1}(t_1) \dots s_{i_{k-1}}(t_{k-1}) \rangle}
{\delta h_{i_k}(t_k)}\right|_{\vec h=0}
\label{multiR0}
\\
R(i_1,t_1,\dots,i_{k-2},t_{k-2}; i_{k-1},t_{k-1},i_{k},t_k) &=& 
\left.
\frac{\delta^2 \langle s_{i_1}(t_1) \dots s_{i_{k-2}}(t_{k-2}) \rangle}
{\delta h_{i_{k-1}}(t_{k-1}) \delta h_{i_k}(t_k)}\right|_{\vec h=0}
\label{multi-R}
\end{eqnarray}
etc.
The angular brackets denote an average over different realizations of the 
dynamics. For a Langevin process as the one in Eq.~(\ref{dyn}) this simply indicates an average over 
thermal noise realizations. The response in Eq.~(\ref{multiR0}) 
corresponds to a case in which the magnetic field $h_{i_k}(t_k)$ couples linearly and 
instantaneously to the $i_k$-th
spin modifying the Hamiltonian at time $t_k$ in such a way that 
$H \to H - h_{i_k}(t_k) s_{i_k}$. Equation~(\ref{multi-R})  defines the 
response of the observable made of a product of $k-2$ spins to kicks applied linearly to two spins $i_{k-1}$ and 
$i_{k}$ at the instants $t_{k-1}$ and $t_k$, respectively. One can easily generalize
this definition to any number of kicks. Note that we do not assume 
any special ordering of times and that the kicked and responsive spins can be the same.
 
When studying models with quenched disorder, as the ones introduced in Section~\ref{dilute}, one is usually interested in their disorder-averaged behavior. Hence one further averages these expressions 
over the probability distribution of disorder and indicates
this calculation by embracing the right-hand-sides with square brackets.
Since under the disorder average 
cross-terms involving different spins typically vanish, one 
usually focuses on the ``global'' correlations and responses. 
In particular, the two-time ones are defined as 
\begin{eqnarray}
C(t_1,t_2) &\equiv& \frac1{N} \sum_{i=1}^N 
\left[ \langle s_{i}(t_1) s_{i}(t_2) \rangle \right]_J
\; ,
\label{def1}
\\
R(t_1,t_2) &\equiv& \frac1{N} \left. \sum_{i=1}^N  
\frac{\delta [\langle s_i(t_1) \rangle]_J}{\delta h_i(t_2)} 
\right|_{\vec h=0} 
\; .
\end{eqnarray}

\subsection{Properties in equilibrium}

In equilibrium the time-dependence of multi-point correlations and 
responses is constrained in several ways. These constraints are a
consequence of the fact that the probability distribution of any 
configuration ${\cal C}$ of a thermostated 
system in equilibrium is given by the time-independent 
Gibbs-Boltzmann distribution $P({\cal C})\propto \exp(-\beta H)$, 
with $\beta$ the inverse temperature and $H$ the Hamiltonian, and that 
causality is expected to hold. (We set the Boltzmann constant $k_B$ to one henceforth.) Moreover, model-independent 
relations between correlations and responses can be established and are the expression of the so-called fluctuation-dissipation theorem ({\sc fdt}).

These equilibrium properties have been worked out in detail 
for two-time functions and they are 
easy to derive using a variety of methods. In particular, for a Langevin
process with white-noise one can exploit its connection to a 
Fokker-Planck equation, or one can use the super-symmetric version 
of the dynamic generating function 
using the symmetries of the action to constrain the properties of the 
observables (see, {\it e.g.}, \cite{yo} for a review). With both 
techniques one finds the following properties of two-time  functions:

\noindent
{\it i.} Time-translation invariance ({\sc tti}), $C(i_1,t_1,i_2,t_2) = 
{\tilde C}(i_1,i_2;t_1-t_2)$, for any two sites and any two-times.

\noindent
{\it ii.} Causality, $R(i_1,t_1;i_2,t_2) = 0$ if $t_2> t_1$.

\noindent
{\it iii.} Fluctuation-dissipation theorem, 
$\beta R(i_1,t_1;i_2,t_2) =  
\partial_{t_2} C(i_1,t_1,i_2,t_2)$, for 
$t_1\geq t_2$.

\noindent
{\it iv.} Reciprocity or Onsager relations, $\langle A(t) B(t') \rangle = 
\langle A(t') B(t) \rangle$, for any pair of observables $A$ and $B$
which are functions of the spins.

For fully-connected models in the thermodynamic 
limit
one can easily prove that all many-point 
correlations decouple into products of two-time ones (if there is an applied magnetic field, a one-time quantity,  the average spin also enters in the decomposition). Hence, 
these quantities have been the focus of most of the analytic
(but also numeric and experimental) studies of glassy systems.

Here, we establish generic properties that multi-point 
correlations and responses must satisfy in equilibrium. We
shall  later 
discuss how these may be generalized for systems slowly 
evolving out of equilibrium. In order to prove these generic properties we use the Fokker-Planck representation of the Langevin equation. 
As special cases we recover the properties {\it i-iv}
of two-time functions.

We focus on a stochastic dynamics of Langevin type
in which each dynamic variable, $s_j$, evolves with
\begin{equation}
\partial_t s_j(t) = -\frac{\delta H}{\delta s_j(t)} + \xi_j(t)
\; .
\label{dyn}
\end{equation}
$\xi_j(t)$ is a Gaussian 
thermal noise with zero mean and white-noise statistics:
\begin{equation}
\langle \xi_j(t)  \xi_k(t')\rangle = 2 T \delta_{jk} \delta(t-t')
\; ,
\label{defnoise}
\end{equation}
with $T$ the temperature of the thermal bath in contact with the system.
We have rescaled time in such a way that the friction coefficient 
is set to one. The time-dependent distribution function $P({\cal C},t)=P(\vec s, t)$ evolves according to  the Fokker-Planck equation:
\begin{equation}
\frac{\partial}{\partial t} P(\vec{s},t) = \sum_j \frac{\partial}{\partial s_j} \left[ 
\frac{\delta H}{\delta s_j(t)} P(\vec{s},t) + T \frac{\partial}{\partial s_j} P(\vec{s},t) \right]
\; .
\end{equation}
In equilibrium this equation is solved by the Gibbs-Boltzmann distribution 
$P_{\rm eq}(\vec{s})=Z^{-1} e^{-\beta H(\vec{s})}$, 
with $Z$ the partition function.
A very useful representation of this equation takes advantage of its similarity with the
Schr\"odinger equation and associates the functions of the stochastic variables with states in quantum mechanics~\cite{KS}
\begin{equation}
f(\vec{s}) \leftrightarrow |f\rangle \; ,\qquad \langle g | f \rangle = \int d\vec{s} \; g(\vec{s})^* f(\vec{s})
\; ,
\end{equation}
and defines position and momentum operators
\begin{eqnarray}
&&
s_j f(\vec{s}) \leftrightarrow \hat{s}_j |f\rangle \; ,
\qquad -i \frac{\partial}{\partial s_j} f(\vec{s}) \leftrightarrow \hat{p}_j |f\rangle\; ,
\qquad [\hat{s}_j , \hat{p}_k] = i \delta_{jk}
\; .
\end{eqnarray}
Note that  $\hat{s}_j^+ = \hat{s}_j$ and
$\hat{p}_j^+ = \hat{p}_j$.
Using this notation the 
Fokker Planck equation reads
\begin{eqnarray*}
\frac{\partial}{\partial t} |P(t)\rangle = H_{\sc fp}  |P(t)\rangle \; ,
\;\;\;
|P(t)\rangle = e^{H_{\sc fp} t} |P(0)\rangle \;,
\;\;\;
H_{\sc fp} = \sum_j \hat{p}_j \left( i H_j(\vec{\hat{s}}) - T \hat{p}_j \right)
\; ,
\end{eqnarray*}
where $H_j \equiv \delta H / \delta s_j$.
One also defines a projection state:
\begin{equation}
\langle - | f \rangle = \int d\vec{s} \; f(\vec{s})
\end{equation}
and writes the equilibrium state as
$|P_{\rm eq}\rangle = Z^{-1} e^{-\beta H(\vec{\hat{s}})} |-\rangle$.
The matrix elements of $\exp(H_{\sc fp}t)$ are transition probabilities. Hence the correlation functions of the observables $A_i$ that are functions of the variables $s_i$
are expressed as 
\begin{equation}
\langle A_n(t_n) \dots A_1(t_1) \rangle = \langle -|\hat{A}_n e^{H_{\sc fp}(t_n-t_{n-1})} \hat{A}_{n-1} \dots \hat{A}_1 e^{H_{\sc fp}t_1} | P(0)\rangle \;, 
\label{generic_correl}
\end{equation}
with $t_n>t_{n-1}>\dots >t_1>0$. $\hat{A}_i$ is obtained from $A_i$ by replacing the variables with the corresponding operators. 
The linear response to a field coupled linearly to the spin corresponds to adding 
$\sum_j -i h_j \hat{p}_j$ to $H_{\sc fp}$. Thus
$\delta/\delta h_j(t) \leftrightarrow -i \hat{p}_j$
in the sense that the effect of the applied field is represented by the insertion of $-i \hat{p}_j$
at time $t$ in a correlator.

Two useful properties also expected are 
time reversal symmetry 
\begin{equation}
e^{\beta H(\vec{\hat{s}})} H_{\sc fp} e^{-\beta H(\vec{\hat{s}})} = H_{\sc fp}^+ \; ,
\label{Hcroix}
\end{equation}
that is due to detailed balance, and 
causality, that implies 
$\langle - | \hat{p}_i =0$, so that the response of any correlator to a field applied after the observation times vanishes.

Let us now describe in detail the extensions of the properties 
{\it i-iv} to multi-time functions. 

\subsubsection{Time translation invariance}

The time-translational invariance of equilibrium correlation functions is transparent in this formalism. Consider Eq.~(\ref{generic_correl}) where all times are shifted by the same  amount $\Delta t$:
\begin{equation}
\langle A_n(t_n + \Delta t) \dots A_1(t_1 + \Delta t) \rangle = \langle -|\hat{A}_n e^{H_{\sc fp}(t_n-t_{n-1})} \hat{A}_{n-1} \dots \hat{A}_1 e^{H_{\sc fp}(t_1 + \Delta t)} | P(0)\rangle \;, 
\label{TTI-0}
\end{equation}
If the system is equilibrated at time $t=0$, $| P(0)\rangle = | P_{\rm eq}\rangle$ and by definition $H_{\sc fp} | P_{\rm eq}\rangle = 0$
so that $\exp(H_{\sc fp}\Delta t) | P_{\rm eq}\rangle = | P_{\rm eq}\rangle$. Hence the time-translational invariance of equilibrium correlation functions:
\begin{equation}
\langle A_n(t_n + \Delta t) \dots A_1(t_1 + \Delta t) \rangle = 
\langle A_n(t_n) \dots A_1(t_1) \rangle 
\; .
\label{TTI}
\end{equation}
Clearly, for a two-time function one recovers property 
{\it i}.

\subsubsection{Generalized Onsager relations}

These express the time reversal symmetry of equilibrium correlation functions. 
Consider $t_n > t_{n-1} > \dots > t_1 =0$ and, 
for simplicity, let us use the spins 
themselves as the observables.  
Then,
\begin{eqnarray}
\langle s_{j_n}(t_n) \dots s_{j_1}(0) \rangle&=&\langle - | \hat{s}_{j_n} e^{H_{\sc fp}(t_n-t_{n-1})} \hat{s}_{j_{n-1}} \dots \hat{s}_{i_2} e^{H_{\sc fp} t_2} \hat{s}_{j_1}|P_{\rm eq}\rangle \\
&=& \langle P_{\rm eq} | \hat{s}_{j_1} e^{H_{\sc fp}^+ t_2} \hat{s}_{j_2} \dots \hat{s}_{j_{n-1}} e^{H_{\sc fp}^+ (t_n-t_{n-1})} \hat{s}_{j_n}| - \rangle \; .
\end{eqnarray}  
Using $\langle P_{\rm eq}| = \langle - | \exp(- \beta H)$, $|-\rangle = \exp( \beta H) | P_{\rm eq} \rangle$ and the relation (\ref{Hcroix}), one obtains:
\begin{eqnarray}
&&
\langle s_{j_n}(t_n) s_{j_{n-1}}(t_{n-1})   \dots s_{j_2}(t_2)  s_{j_1}(0) \rangle = 
\nonumber\\
&&
\;\;\;\;\;\;\;\;\;\;\;\;\;\;\;\;
\langle s_{j_1}(t_n) s_{j_2}(t_n - t_2)  \dots s_{j_{n-1}}(t_n-t_{n-1}) s_{j_n}(0) \rangle
\; .
\end{eqnarray}

The usual Onsager relation on two-point functions is a particular case and reads:
 $\langle s_j(t) s_k(0) \rangle = \langle s_k(t) s_j(0) \rangle$.
For a three-time correlator  one has  
$\langle s_l(-t_2) s_k(t_1) s_j(t_2) \rangle $$= $$\langle s_j(-t_2) s_k(- t_1) s_l(t_2) \rangle$, for $t_2 > |t_1|$. If $l=j$, with the two extreme times fixed, the correlation is an even function of $t_1$.

Similar relations can be derived for generic functions of the spins
 as long as they do not include, when  expressed in the 
quantum mechanical language, the operator $\hat p$. In the 
two-time case these reduce to property {\it iv}. 

\subsubsection{{\sc fd} relation on two-time functions}

According to the previous definitions, the usual equilibrium linear response function reads:
\begin{equation}
R_{jk}(t)=\left.
\frac{\delta}{\delta h_k(0)} \langle s_j(t) \rangle \right|_{h=0}= \langle -| \hat{s}_j e^{H_{\sc fp}t} (- i \hat{p}_k ) | P_{\rm eq} \rangle
\; , 
\end{equation}
where we simplify again the presentation by considering the 
response of the simple observable given by a spin to a 
perturbation done on another spin. Generalizations to more 
complicated observables are straightforward.

On the other hand, the equilibrium correlation function and its time derivative are
\begin{eqnarray}
C_{jk}(t)&=&\langle s_j(t) s_k(0) \rangle = \langle - | \hat{s}_j e^{H_{\sc fp} t} \hat{s}_k | P_{\rm eq} \rangle 
\; ,
\\
\frac{d}{dt} C_{jk}(t)&=& \langle - | \hat{s}_j e^{H_{\sc fp} t} H_{\sc fp} \hat{s}_k | P_{\rm eq} \rangle
\; .
\end{eqnarray}
By definition of the equilibrium probability distribution, $(T i \hat{p}_k + H_k(\hat{\vec{s}}))|P_{\rm eq}\rangle = 0 \quad \forall k$. Using this relation and the commutation properties of the operators, one obtains
\begin{equation}
H_{\sc fp} \hat{s}_k|P_{eq}\rangle = T i \hat{p}_k |P_{\rm eq}\rangle
\label{HsigsurPeq}
\end{equation}
and thus recovers the usual {\sc fd} relation in equilibrium:
\begin{equation}
R_{jk}(t) = - \frac{1}{T} \frac{d}{d t} C_{jk}(t) \; , \qquad t>0
\label{UsualFD}
\end{equation}

\subsubsection{{\sc fd} relation on 3 point functions}
\label{FDT3}

\noindent
{\it Response to a single kick}\\

The first type of three point response function one has to investigate is the response of a two point correlator to a single kick. It is clear by causality that the response vanishes if the kick is posterior to the observation times. Two cases must then be distinguished:
\begin{itemize}
\item The perturbation is done prior to the two observation times. Let us choose $t_2 > t_1 > t_0$,
\begin{equation}
\frac{\delta}{\delta h_j(t_0)} \langle s_k(t_1) s_l(t_2) \rangle = 
\langle -| \hat{s}_l e^{H_{\sc fp}(t_2-t_1)} \hat{s}_k e^{H_{\sc fp}(t_1-t_0)}  (- i \hat{p}_j ) | P_{\rm eq} \rangle
\; .
\end{equation}
The three time correlator and its derivative with respect to the earlier time are:
\begin{eqnarray}
\langle s_j(t_0) s_k(t_1) s_l(t_2) \rangle&=&\langle -| \hat{s}_l e^{H_{\sc fp}(t_2-t_1)} \hat{s}_k e^{H_{\sc fp}(t_1-t_0)} \hat{s}_j  | P_{\rm eq} \rangle 
\; ,\\
\frac{\partial}{\partial t_0} \langle s_j(t_0) s_k(t_1) s_l(t_2) \rangle&=&- \langle -| \hat{s}_l e^{H_{\sc fp}(t_2-t_1)} \hat{s}_k e^{H_{\sc fp}(t_1-t_0)} H_{\sc fp} \hat{s}_j  | P_{\rm eq} \rangle
\; .
\end{eqnarray}
Using (\ref{HsigsurPeq}), one obtains:
\begin{equation}
\frac{\delta}{\delta h_j(t_0)} \langle s_k(t_1) s_l(t_2) \rangle = \frac{1}{T} \frac{\partial}{\partial t_0} \langle s_j(t_0) s_k(t_1) s_l(t_2) \rangle
\; .
\label{simple-FDT}
\end{equation}
This relation is a natural generalization of the usual {\sc fdt}.

\item The perturbation time is in between the two observation times. For the sake of clarity let us denote the three times $t_2 > t_1 > -t_2$. By time translational invariance, we do not loose any generality with this choice. Then,
\begin{equation}
\frac{\delta}{\delta h_k(t_1)} \langle s_j(-t_2) s_l(t_2) \rangle =
\langle - | \hat{s}_l e^{H_{\sc fp}(t_2-t_1)} (- i \hat{p}_k ) e^{H_{\sc fp}(t_1 + t_2)} \hat{s}_j |P_{\rm eq}\rangle
\; .
\end{equation}
Considering the time-reversed (conjugated) expression, with $j$ and $l$ exchanged and $t_1$ reversed, and using the Baker-Campbell-Hausdorff formula $e^{-\beta H} \; $$\hat{p}_k$$ e^{\beta H} = \hat{p}_k - i \beta H_k$, this can be rewritten as
\begin{equation}
\frac{\delta}{\delta h_k(-t_1)} \langle s_l(-t_2) s_j(t_2) \rangle =
\langle - | \hat{s}_l e^{H_{\sc fp}(t_2-t_1)} (i \hat{p}_k + \beta H_k ) e^{H_{\sc fp}(t_2 + t_1)} \hat{s}_j |P_{\rm eq}\rangle
\end{equation}
Consider now the derivative with respects to the intermediate time of the three time correlator:
\begin{equation}
\frac{\partial}{\partial t_1} \langle s_j(-t_2) s_k(t_1) s_l(t_2) \rangle = \langle - | \hat{s}_l e^{H_{\sc fp}(t_2-t_1)} [\hat{s}_k,H_{\sc fp}] e^{H_{\sc fp}(t_1 + t_2)} \hat{s}_j |P_{\rm eq}\rangle
\; .
\end{equation}
Since $[\hat{s}_k,H_{\sc fp}]=-2 i T \hat{p}_k - H_k$, we have the following {\it model independent} {\sc fd} relation:
\begin{eqnarray}
\frac{1}{T} \frac{\partial}{\partial t_1} \langle s_j(-t_2) s_k(t_1) s_l(t_2) \rangle &=& \frac{\delta}{\delta h_k(t_1)} \langle s_j(-t_2) s_l(t_2) \rangle 
\nonumber\\
&& - \frac{\delta}{\delta h_k(-t_1)} \langle s_l(-t_2) s_j(t_2) \rangle
\; .
\label{two-FDT}
\end{eqnarray}
\end{itemize}
Note that while on the left-hand-side we have the ``expected''
variation of the three-time correlator, on the right-hand-side two responses appear (and none vanishes due to causality). 

\noindent \\
{\it Response to two kicks}\\

One can also construct the response of an observable to two earlier kicks. Taking $t_2 > t_1 > t_0$,
\begin{equation}
\frac{\delta^2 \langle s_l(t_2) \rangle}{\delta h_j(t_0) \delta h_k(t_1)}  = \langle - | \hat{s}_l e^{H_{\sc fp}(t_2-t_1)} (-i \hat{p}_k) e^{H_{\sc fp}(t_1 - t_0)} (-i \hat{p}_j) |P_{eq}\rangle
\; ,
\end{equation}
and using Eq.~(\ref{HsigsurPeq}) one obtains
\begin{eqnarray}
\frac{\delta^2 \langle s_l(t_2) \rangle}{\delta h_j(t_0) \delta h_k(t_1)}  &=& \frac{1}{T} \frac{\partial}{\partial t_0} \left( \frac{\delta}{\delta h_k(t_1)} \langle s_j(t_0) s_l(t_2) \rangle \right) 
\nonumber\\
&=& \frac{1}{T^2} \frac{\partial^2}{\partial t_0 \partial t_1} \langle s_j(t_0) s_k(t_1) s_l(t_2) \rangle 
\nonumber\\
&&
+\frac{1}{T} \frac{\partial}{\partial t_0} \left( \frac{\delta}{\delta h_k(t_0+t_2-t_1)} \langle s_l(t_0) s_j(t_2) \rangle \right)
\label{two-kicks}
\end{eqnarray}

Interestingly enough, 
we find that the three point correlator does not determine completely the three point responses. 

\subsection{A general fluctuation principle}
\label{FluctuationSection}
\def\straj{\underline{s}}
\def\htraj{\underline{h}}
\def\xtraj{\underline{x}}
\def\phitraj{\underline{\phi}}
\newcommand{\<}{\langle}
\renewcommand{\>}{\rangle}

In the previous Section we derived some {\sc fd} relations on a 
case-by-case basis.
One may wonder whether these relations can be expressed in a unified way. 

For the sake of simplicity, let us consider the case of a
single time-varying field $h(t)$, coupled linearly to any given 
observable ${\cal O}(s)$ (here we denote by $s$ the whole spin
configuration: $s\equiv\{s_1\dots s_N\}$).
In order to state a unifying principle for the multi-time 
{\sc fd} relations, we need some quantity which encodes the
full hierarchy of correlation and response functions
defined above.  
Let us denote by $\xtraj_{-t_M,t_M}$ a particular trajectory of the observable
$x$ between the times $-t_M$ and $t_M$. Hereafter we shall assume $t_M$ to be larger
than any other time in the problem and we shall drop the subscripts. 
Consider the probability density $dP(\straj|\htraj)$ 
of a trajectory $\straj$ of the system,
given a particular realization $\htraj$ of the external field.
It is clear that, by integrating $dP(\straj|\htraj)$  over $\straj$ 
and expanding in powers of $\htraj$, we can recover all the correlation
and response functions. 

In order to state a fluctuation principle for $P(\straj|\htraj)$,
we need to define the time reversal operation. The time reversed 
of $\xtraj$ is denoted by $\xtraj^R$ and is defined by 
$x^R(t) \equiv x(-t)$. 
We make the following assumptions: $(i)$ the 
probability distribution at the initial time $-t_M$ is the equilibrium one 
in zero-field: $P_{\rm eq}(s)$; $(ii)$ the dynamics satisfies detailed balance;
$(iii)$ the perturbing
field vanishes outside the time interval $[-t_M,t_M]$.
Under these hypotheses, it is easy to show that
\begin{eqnarray}
\frac{dP(\straj^R|\htraj^R)}{dP(\straj|\htraj)} = \exp\left\{ -\beta\int_{-t_M}^{t_M}
\!\!dt\,h(t)\,\dot{\cal O}(t)\right\}\, , \label{FluctuationPrinciple}
\end{eqnarray}
where $\dot{\cal O}(t)$ denotes the time derivative of the observable 
${\cal O}(s)$ along the trajectory $\straj$.

The proof of Eq. (\ref{FluctuationPrinciple}) is 
straightforward\footnote{Nevertheless the result (\ref{FluctuationPrinciple})
holds under more general hypotheses, for instance in a discrete-time 
Markov chain.} if we assume Ito discretization of the 
Langevin dynamics (\ref{dyn}). In this case the probability density
of a trajectory $\straj$ can be written explicitly
\begin{eqnarray}
dP(\straj|\htraj) = P_{\rm eq}(s(-t_M))\left\<\prod_{i,t}\delta\left(
\xi_i(t)-\dot{s}_i(t)-\frac{\delta H}{\delta s_i}+h(t)
\frac{\delta {\cal O}}{\delta s_i}\right)\right\>_{\xi}\cdot d\straj
\; ,
\end{eqnarray}
where $P_{\rm eq}(s)$ is the equilibrium distribution. 
Using this expression one obtains
\begin{eqnarray}
\frac{dP(\straj^R|\htraj^R)}{dP(\straj|\htraj)} = 
\frac{P_{\rm eq}(s(t_M))}{P_{\rm eq}(s(-t_M))}\exp\left\{ -\beta\sum_i\int_{-t_M}^{t_M}
\!\!dt\,\left[-\frac{\delta H}{\delta s_i}+h(t)
\frac{\delta {\cal O}}{\delta s_i}\right]\dot{s}_i(t)\right\}\, ,
\end{eqnarray}
which reduces to (\ref{FluctuationPrinciple}) upon the insertion of the
Boltzmann distribution $P_{\rm eq}(s)\propto \exp(-\beta H(s))$.

It can be useful to formulate a few simple remarks on this result:
\begin{itemize}
\item In the $\htraj = 0$ case, Eq.~(\ref{FluctuationPrinciple}) is simply a 
rephrasing of time-reversal invariance. A time-varying external field 
violates this invariance. The amount of such a violation is quantified 
by the work done by the field on the system.
\item Equation (\ref{FluctuationPrinciple}) can be regarded as the dynamic 
analogous of the following identity
\begin{eqnarray}
\frac{dP_{\rm eq}(s|-h)}{dP_{\rm eq}(s|h)} = 
\exp\left\{-2\beta h {\cal O}(s)\right\}\, ,
\end{eqnarray}
which holds if $dP_{\rm eq}(s|-h)$ is the probability of the configuration $s$
according to the Boltzmann distribution.
\item Using Eq.~(\ref{FluctuationPrinciple}) we can 
recover the {\sc fd} relations of the
previous Section. As a simple exercise, let us consider the 
relation between two-point functions (\ref{UsualFD}). We take
${\cal O}(s)=s_k$, multiply both sides of Eq.~(\ref{FluctuationPrinciple})
by $s_j(t_1)dP(\straj|\htraj)$ and integrate over $\straj$. This yields
\begin{eqnarray}
\<s_j(-t_1)\>_{\htraj^R} = \< s_j(t_1)\cdot
\exp\left\{-\beta\int\!dt'\, h_k(t')\dot{s}_k(t')\right\}\>_{\htraj}\, ,
\label{NonZeroField}
\end{eqnarray}
where $\<\cdot\>_{\htraj}$ denotes the average under applied magnetic field
$\htraj$.
Expanding the two members of this identity in powers of $\htraj$, we get,
at the linear order:
\begin{eqnarray}
\frac{1}{T}\frac{\partial}{\partial t_2} 
\<s_j(t_1)s_k(t_2)\> = \frac{\delta\<s_j(t_1)\>}{\delta h_k(t_2)}-
\frac{\delta\<s_j(-t_1)\>}{\delta h_k(-t_2)} \; ,
\end{eqnarray}
which, using causality yields back Eq.~(\ref{UsualFD}).

The general strategy for deriving {\sc fd} relations for multi-time
functions is easily stated: $(i)$ multiply both sides of 
Eq.~(\ref{FluctuationPrinciple}) by the quantity 
$s_{i_1}(t_1)\cdots s_{i_m}(t_m)\cdot dP(\straj|\htraj)$; 
$(ii)$ integrate over 
$\straj$; $(iii)$ expand in powers of $\htraj$ and collect the terms
multiplying $h(t_{m+1})\cdots h(t_{m+n})$. This procedure yields 
a relation for $(m+n)$-times functions. Of course if we want 
to study the response to kicks on several distinct spins
we must consider the obvious generalization of Eq. (\ref{FluctuationPrinciple})
to the case of several observables.
\item Equation (\ref{FluctuationPrinciple}) can be used to derive
relations which are exact with a non-vanishing perturbing field.
A simple example is Eq. (\ref{NonZeroField}). With an appropriate
choice of the time dependence of $h_k(t)$, this result is amenable
for a numerical check. Unlike for the usual
{\sc fd} theorem (\ref{UsualFD}), one is not obliged
to take the zero-field limit which can be numerically tricky.
\item Finally, the principle (\ref{FluctuationPrinciple}) is quite
reminiscent of the Gallavotti-Cohen ({\sc gc}) theorem~\cite{GC,GC2} as stated in Ref.~\cite{Jorge-GC} (see also ~\cite{Spohn,Sellitto})
for stochastic dynamics. However {\sc gc}  refers to stationary systems
and has a non-trivial content only if the dynamics violates the detailed 
balance. In Eq.~(\ref{FluctuationPrinciple}) we consider the 
complementary situation. Detailed balance is satisfied at any time, but 
time-reversal invariance is violated by the explicit time-dependence of the 
external field. Hopefully this type of result is more suitable for the present
context.
\end{itemize}

\subsection{Extensions out of equilibrium}

The non-equilibrium slow dynamics of glassy systems
presents rather simple modifications of the equilibrium properties of two-time correlations and 
responses~\cite{Cuku1,Cuku2,yo}. 
The two-time global correlations decay (increase) monotonically as a function of the longer (shorter) time.
This property allows one to propose ``triangular relations'' that link, 
in the limit of long times, any two-time correlation to other two
evaluated at an intermediate time:~\cite{Cuku2}
\begin{equation}
C(t_1,t_3) = f(C(t_1,t_2), C(t_2,t_3))
\;\;\;\;\;\;\;\; t_1 \geq t_2 \geq t_3
\; .
\end{equation}
Using very general arguments one then proves that 
within a correlation scale (see~\cite{Cuku2} for its precise definition)
the global two-time correlator behaves as 
\begin{equation}
C(t_1,t_2) \equiv \frac1{N} \sum_{i=1}^N 
\left[ \langle s_{i}(t_1) s_{i}(t_2) \rangle \right]_J
= f\left( \frac{l(t_2)}{l(t_1)}\right)
\end{equation}
with $l(t)$ a monotonic growing function. Moreover,
in the limit of long times one finds that  
the {\sc fdt} is modified to 
\begin{equation}
R(t_1,t_2) \equiv \frac1{N} \left. \sum_{i=1}^N  
\frac{\delta [\langle s_i(t_1) \rangle]_J}{\delta h_i(t_2)} 
\right|_{\vec h=0} 
=\theta(t_1-t_2) \frac{1}{T_{\sc eff}} \frac{\partial C(t_1,t_2)}{\partial t_2}
\; ,
\end{equation}
with $T_{\sc eff}$ a correlation-scale dependent effective temperature~\cite{Cukupe}. In this Section we discuss 
possible generalizations of these properties to the 
case of multi-time functions.

\subsubsection{Time-scalings}

Let us first discuss the consequences of having multiple correlation 
scales, as defined via the behavior of the two-time global correlator,
on the multi-time {\it global} ones. The latter 
are defined from Eqs.~(\ref{def1}):
\begin{equation}
C(t_1,t_2,\dots, t_n) \equiv \frac1{N} \sum_{i} C(i,t_1,i,t_2,\dots, i,t_n)
\; ,
\end{equation}
and are the ones expected to be relevant in a disorder
average treatment.

  For concreteness, take a system 
with two correlation scales (as happens for the fully connected $p$ spin 
model and, presumably, for the dilute case too). This means that 
in the asymptotic limit in which $t_1\geq t_2$ are both long but 
their ratio $t_1/t_2$ vary between one and infinity the times are classified
according to the value of $C(t_1,t_2)$.
If $t_1$ and $t_2$ are near by and 
$C(t_1,t_2) \geq q_{\sc ea}$, we are in the fast correlation scale.
If $t_1$ and $t_2$ are far away, 
$C(t_1,t_2) < q_{\sc ea}$, and we are in the slow correlation scale.

Taking three times $t_1\geq t_2\geq t_3$ we have four possibilities:

\noindent
{\it i.} The three times are nearby with all 
correlations being larger than $q_{\sc ea}$. 

\noindent
{\it ii.}
The longer times $t_1$ and $t_2$ 
are nearby while the shortest one $t_3$ is far away; in this case,
$C(t_1,t_2) \geq q_{\sc ea} \geq C(t_1,t_3)$ and 
$C(t_1,t_2) \geq q_{\sc ea} \geq C(t_2,t_3)$. 

\noindent
{\it iii.} The 
``reversed'' situation in which the two shorter times $t_2$ and $t_3$ are
nearby and the longest one $t_1$ is far away; then
$C(t_1,t_2) \leq q_{\sc ea} \leq C(t_2,t_3)$ and 
$C(t_1,t_3) \leq q_{\sc ea} \leq C(t_2,t_3)$. 

\noindent
{\it iv.}
Finally the 
three times can be far away from each other in which case all 
correlations are smaller than $q_{\sc ea}$.

The property of monotonicity can be used to express any muti-point 
correlator in terms of two-time ones. Take for instance a correlator 
evaluated on three times, $C(t_1,t_2,t_3)$. Using the monotonicity property one can 
invert the two-time correlation $C_{12} \equiv C(t_1,t_2)$ and  write $t_1=g(C_{12},t_2)$.
Equivalently, $t_2=g(C_{23},t_3)$. Thus, 
\begin{eqnarray}
C(t_1,t_2,t_3) &=& C(g(C_{12}, g(C_{23},t_3)), g(C_{23},t_3), t_3) 
\; .
\end{eqnarray}
If we assume that the limit $t_1\geq t_2\geq t_3\to\infty$ while  $C_{12}$ and $C_{23}$ are
kept fixed exists then 
\begin{eqnarray}
C(t_1,t_2,t_3) &=& f_1^{(3)}(C_{12}, C_{23})
\; .
\end{eqnarray}
Clearly, we could have chosen to work with $t_1=g(C_{13},t_3)$ and 
obtain 
\begin{eqnarray}
C(t_1,t_2,t_3) &=& f_2^{(3)}(C_{13}, C_{23})
\; .
\end{eqnarray}
The liberty to choose representation is clear in the equilibrium case where one 
can write
\begin{eqnarray*}
C(t_1,t_2,t_3) = f_{{\rm eq},1}^{(3)}(t_1-t_2, t_2-t_3) = f_{{\rm eq},2}^{(3)}(t_1-t_3, t_2-t_3) =
f_{{\rm eq},3}^{(3)}(t_1-t_3, t_1-t_2) 
\; .
\end{eqnarray*}
With similar arguments one can write any multi-time correlation in terms of 
two-time correlations only admitting that the limits exist. In all cases we 
choose to take the limit of the shortest time to infinity to ensure that all 
two-time correlations are in their asymptotic regime. 

As an example consider a case in which there are two correlation scales
and that the three times are chosen in such a way that two of them, $t_1$ and $t_2$,
are nearby
so that the two-time correlation $C(t_1,t_2)$  falls above $q_{\sc ea}$ and the third one is far away
from both in such a way that the two-time correlations $C(t_1,t_3)$ and $C(t_2,t_3)$ 
fall below $q_{\sc ea}$. Then we expect 
\begin{eqnarray*}
C(t_1,t_2,t_3) = \tilde f\left(t_1-t_2, \frac{l(t_2)}{l(t_3)} \right) 
\; .
\end{eqnarray*}
This relation (and similar ones) generalize time-translational invariance for multi-time 
correlations  [see Eq.~(\ref{TTI})] to the slowly evolving non-equilibrium case.

\subsubsection{Fluctuation -- dissipation relations}
Another important feature of glassy dynamics is the modification of 
the fluctuation -- dissipation theorem. 
The corresponding out-of-equilibrum fluctuation--dissipation
relation ({\sc ofdr}) for two-point functions has been the object of 
intensive studies in the last years (see~\cite{yo} for a review).
In Section~\ref{FDT3} we obtained fluctuation -- dissipation 
relations for multi-time correlations and responses. 
Here, we discuss the possible form of the corresponding multi-time {\sc ofdr}'s. 

Guessing the correct generalization of the multi-time
{\sc fd} relations is quite difficult. Already in equilibrium,
the form of these relations is far from obvious and a careful derivation 
was necessary. Here we shall adopt the following approach.
We consider the multi-point correlation and response functions of
a Gaussian model. In this case the two-time {\sc ofdr} straightforwardly 
implies multi-time {\sc ofdr}'s. We then rewrite these relations
in a model-independent fashion. This can be done in a particular
compact way by modifying the fluctuation principle of Sec.
\ref{FluctuationSection}.

For the sake of simplicity, we shall consider the case of 
a scalar field $\phi(t)$ with $t\ge 0$, linearly coupled to an
external field $h(t)$. Being Gaussian, its behavior is completely
specified by the following quantities:
\begin{eqnarray}
M(t) = \<\phi(t)\> \, , \;\; C(t,t') = \<\phi(t)\phi(t')\> 
\, , \;\;R(t;t') = \frac{\delta\<\phi(t)\>}{\delta h(t')} \, .
\label{FunctionsDefinition}
\end{eqnarray}
As a warm-up exercise, let us consider three point functions. It is 
simple to show that, in the Gaussian case:
\begin{eqnarray}
C(t_1,t_2,t_3) & = & M(t_1)M(t_2)M(t_3)+M(t_1)C(t_2,t_3) + M(t_2)C(t_1,t_3)+
\nonumber\\
&& +M(t_3)C(t_1,t_2)\, ,\\
R(t_2,t_3;t_1) & = & M(t_2)R(t_2;t_1)+M(t_2)R(t_3;t_1)\, ,\\
R(t_3;t_2,t_1) & = & 0\, ,
\end{eqnarray}
where we chose the ordering of times $t_1 < t_2 < t_3$, which we shall keep in this Section.
We should now make some assumptions on the long time
behavior of the functions in Eq. (\ref{FunctionsDefinition}). To keep
the presentation as simple as possible we shall consider a scenario with two correlation scales:
\begin{eqnarray}
M(t) & \approx & M_{\rm eq}\, ,\\
C(t,t') & \approx & C_{\rm eq}(t-t') + C_{\rm ag}(l(t)/l(t'))\, ,\\
R(t;t') & \approx & R_{\rm eq}(t-t') + {\tilde l}(t') R_{\rm ag}(l(t)/l(t'))\, .
\end{eqnarray}
with 
\begin{eqnarray}
\beta\,\partial_{\tau} C_{\rm eq}(\tau) = -R_{\rm eq}(\tau) \, ,\;\;\;
\beta_{\sc eff} \, \partial_{t'} C_{\rm ag}(l(t)/l(t')) = {\tilde l}(t') R_{\rm ag}(l(t)/l(t'))\, .\label{FdtOfdr}
\end{eqnarray}
We defined ${\tilde l}(t)=l'(t)/l(t)$ that we assume to vanish when $t \to \infty$. All the times in our discussion below are such that the above asymptotic
forms are well verified. 

Let us consider
the different cases for the response to a single kick (we leave 
the two-kick case as an exercise for the reader):
\begin{itemize}
\item The perturbation is done at time $t_1$, i.e. 
prior to the two observation times.
We must distinguish several different possibilities according to the scaling
of the time separations $(t_2-t_1)$ and $(t_3-t_2)$, as 
$t_1,t_2,t_3\to\infty$:
\begin{itemize}
\item $(t_2-t_1)$ and $(t_3-t_2)$ of $O(1)$.  We get 
\begin{eqnarray}
\partial_{t_1}C(t_1,t_2,t_3) & \approx & 
-M_{\rm eq}\partial C_{\rm eq}(t_3-t_1)-M_{\rm eq}\partial
C_{\rm eq}(t_2-t_1)\, ,\label{Eq1}\\
R(t_2,t_3;t_1) & \approx & M_{\rm eq} R_{\rm eq}(t_3-t_1)+
M_{\rm eq} R_{\rm eq}(t_2-t_1)\, ,\label{Eq1_2}
\end{eqnarray}
where we noted by $\partial C(\cdot)$ the derivative of $C(\cdot)$ 
with respect to its argument.
From Eqs.~(\ref{Eq1}) and (\ref{Eq1_2}) we recover the {\sc fd} relation 
for three time functions:
\begin{eqnarray}
\beta\, \partial_{t_1}C(t_1,t_2,t_3) & \approx & R(t_2,t_3;t_1)\, .
\label{fdt3}
\end{eqnarray}
\item $(t_2-t_1)=O(1)$,  $(t_3-t_2) \to\infty$. In this case we have
\begin{eqnarray}
\partial_{t_1}C(t_1,t_2,t_3) & \approx & 
-M_{\rm eq}\frac{l(t_3)l'(t_1)}{l(t_1)^2}\partial C_{\rm ag}(l(t_3)/l(t_1))
-M_{\rm eq}\partial C_{\rm eq}(t_2-t_1)\, ,\label{Oeq1}\\
R(t_2,t_3;t_1) & \approx & M_{\rm eq} \, {\tilde l}(t_1) \,R_{\rm ag}(l(t_3)/l(t_1))+
M_{\rm eq} R_{\rm eq}(t_2-t_1)\, .\label{Oeq2}
\end{eqnarray}
At first sight it may seem that Eqs.~(\ref{Oeq1}) and (\ref{Oeq2})
are much more difficult to deal with than Eqs.~(\ref{Eq1}) and (\ref{Eq1_2}).
Notice however than the first terms in the above expressions
are of order ${\tilde l}(t_1)$ with respect to the second ones, and can therefore 
be dropped in the aging limit. In this limit the {\sc fd} relation
(\ref{fdt3}) is once again saisfied.
\item $(t_2-t_1)\to \infty$,  $(t_3-t_2)= O(1)$. We have
\begin{eqnarray}
\partial_{t_1}C(t_1,t_2,t_3) & \approx & 
M_{\rm eq}\partial_{t_1} C_{\rm ag}(l(t_3)/l(t_1))
+M_{\rm eq}\partial_{t_1} 
C_{\rm ag}(l(t_2)/l(t_1)) \, ,\\
R(t_2,t_3;t_1) & \approx & M_{\rm eq} {\tilde l}(t_1) \, R_{\rm ag}(l(t_3)/l(t_1))+
M_{\rm eq} {\tilde l}(t_1) \, R_{\rm ag}(l(t_2)/l(t_1)) \, .
\label{Eq2}
\end{eqnarray}
Using Eq.~(\ref{FdtOfdr}), it is easy to show that these functions
satisfy the natural out-of-equilibrium generalization of Eq.~(\ref{fdt3}):
\begin{eqnarray}
\beta_{\sc eff}\, \partial_{t_1}C(t_1,t_2,t_3) & \approx & R(t_2,t_3;t_1) \, .
\label{ofdr3}
\end{eqnarray}
\item $(t_2-t_1)\to\infty$,  $(t_3-t_2) \to \infty$. This case is similar to 
the previous one. The relation (\ref{ofdr3}) is recovered.
\end{itemize}
\item The perturbation time is $t_2$, i.e. in between the two 
observation times.
This case is more interesting than the previous one.
As before, we need to treat separately the different time-scalings:
\begin{itemize}
\item $(t_2-t_1)$, $(t_3-t_2) = O(1)$.  This case
is quite simple:
\begin{eqnarray}
\partial_{t_2}C(t_1,t_2,t_3) & \approx & 
-M_{\rm eq}\partial C_{\rm eq}(t_3-t_2)+M_{\rm eq}\partial
C_{\rm eq}(t_2-t_1)\, ,\\
R(t_1,t_3;t_2) & \approx &
M_{\rm eq} R_{\rm eq}(t_3-t_2)\, ,\\
R(t^R_1,t^R_3;t^R_2) & \approx &
M_{\rm eq} R_{\rm eq}(t_2-t_1)\, ,
\end{eqnarray}
where we defined the time-reversal operation as
follows
\begin{eqnarray}
t\mapsto t^R \equiv t^*-t
\end{eqnarray}
for some (large) fixed time $t^*$.
Using Eq.~(\ref{FdtOfdr}) we re-obtain the equilibrium relation,
{\it cfr.}. Eq.~(\ref{two-FDT})
\begin{eqnarray}
\beta\partial_{t_2} C(t_1,t_2,t_3) \approx 
R(t_1,t_3;t_2)-R(t^R_1,t^R_3;t^R_2)\, .
\label{fdt3_2}
\end{eqnarray}
\item $(t_2-t_1)=O(1)$,  $(t_3-t_2) \to\infty$. We have
\begin{eqnarray}
\partial_{t_2}C(t_1,t_2,t_3) & \approx & 
-M_{\rm eq}\frac{l(t_3)l'(t_2)}{l(t_2)^2}
\partial C_{\rm ag}(l(t_3)/l(t_2))+M_{\rm eq}\partial C_{\rm eq}(t_2-t_1)\, ,
\label{Correlation_1}\\
R(t_1,t_3;t_2) & \approx &
M_{\rm eq} {\tilde l}(t_2)R_{\rm ag}(l(t_3)/l(t_2))\, ,\label{Response1_1}\\
R(t^R_1,t^R_3;t^R_2) & \approx &
M_{\rm eq} R_{\rm eq}(t_2-t_1)\, .
\end{eqnarray}
Notice that the first contribution to the correlation, {\it cfr.} 
Eq.~(\ref{Correlation_1}),  and the response (\ref{Response1_1}) are of order 
${\tilde l}(t_2)$ with respect to the other terms, and therefore vanish in the 
aging limit. One recovers therefore the equilibrium relation 
(\ref{fdt3_2}).
\item $(t_2-t_1)\to\infty$,  $(t_3-t_2) = O(1)$. This case 
works exactly as the previous one: the equilibrium {\sc fd} relation
(\ref{fdt3_2}) holds up to terms of relative order ${\tilde l}(t_1)$.
\item $(t_2-t_1)$,  $(t_3-t_2) \to\infty$. Here something interesting 
finally happens:
\begin{eqnarray}
\partial_{t_2}C(t_1,t_2,t_3) & \approx & 
M_{\rm eq}\partial_{t_2} C_{\rm ag}(l(t_3)/l(t_2))+
M_{\rm eq}\partial_{t_2} C_{\rm ag}(l(t_2)/l(t_1))\, ,
\label{Correlation}\\
R(t_1,t_3;t_2) & \approx &
M_{\rm eq} {\tilde l}(t_2)R_{\rm ag}(l(t_3)/l(t_2))\, ,\label{Response1}\\
R(t^{Rl}_1,t^{Rl}_3;t^{Rl}_2) & \approx &
M_{\rm eq} {\tilde l}(t_1)R_{\rm ag}(l(t_2)/l(t_1))\, ,
\end{eqnarray}
where we defined the ``aging'' time-reversal transformation as follows:
\begin{eqnarray}
t\mapsto t^{Rl} \equiv l^{-1}\left(\frac{l(t^*)^2}{l(t)}\right)\, ,
\label{aging_time_reversal}
\end{eqnarray}
for some (large) fixed time $t^*$. Using this expression 
and the relations (\ref{FdtOfdr}) one can easily show that
\begin{eqnarray}
\beta_{\sc eff}\partial_{t_2} C(t_1,t_2,t_3) \approx 
R(t_1,t_3;t_2)-R(t^{Rl}_1,t^{Rl}_3;t^{Rl}_2)\, .
\end{eqnarray}
In other words, the out-of-equilibrium {\sc fd} relation is obtained
from the equilibrium one by replacing the thermodynamic 
temperature with an effective one, and by a {\it redefinition of the 
time-reversal operation}.
\end{itemize}
\end{itemize}

At this point it is easy to summarize the above results
(and the similar ones which can be derived for higher correlations)
along the lines of Sec \ref{FluctuationSection}.
More precisely, let us split the field in its fast components 
plus the slowly varying part~\cite{Zannetti,SilvioMiguel}:
\begin{eqnarray}
\phi(t) = \phi_{\rm eq}(t) + \phi_{\rm ag}(t)\, ,\;\;\;
h(t) = h_{\rm eq}(t) + h_{\rm ag}(t)\, .
\end{eqnarray}
While $\phi_{\rm eq}(t)$ obeys a fluctuation principle of the form
(\ref{FluctuationPrinciple}), this has to be modified as follows for
$\phi_{\rm ag}(t)$:
\begin{eqnarray}
\frac{dP(\phitraj_{\rm ag}^{Rl}|\htraj_{\rm ag}^{Rl})}{dP(\phitraj_{\rm ag}|\htraj_{\rm ag})} \approx \exp\left\{ -\beta_{\sc eff}\int
\!\!dt\,h_{\rm ag}(t)\,\dot{\phi}_{\rm ag}(t)\right\}\, .\label{GeneralizedFluctuationPrinciple}
\end{eqnarray}
Here $\htraj^{Rl}$ and $\phitraj^{Rl}$ are defined analogously to
$\htraj^{R}$ and $\phitraj^{R}$ ({\it cfr.} Eq.~(\ref{FluctuationPrinciple})) 
but using the 
time-reversal operation (\ref{aging_time_reversal}) instead of the usual one.
The reader can easily check Eq.~(\ref{GeneralizedFluctuationPrinciple})
on a Gaussian process. It is natural to conjecture it to hold even for 
non-Gaussian ones. It would be of great interest to check this 
conjecture in a numerical simulation. 

Let us by the way notice that Eq. (\ref{GeneralizedFluctuationPrinciple})
implies a whole class of out-of-equilibrium Onsager relations. A simple 
example is
\begin{eqnarray}
C_{\rm ag}(t_1,t_2,t_3) = C_{\rm ag}(t_1,t_2',t_3)\;\;\;\;\;\;\;
\mbox{if}\;\;\;\;    l(t_2)l(t_2')=l(t_1)l(t_3)\, .
\end{eqnarray}

\section{Dilute disordered spin systems}
\label{dilute}

Dilute disorder spin models are typical cases where a description 
in terms of two-point functions only is not complete. Multi-point 
correlations do not factorize and carry non-trivial information 
about the dynamics of these systems.

In this Section we introduce a method to solve the disorder averaged dynamics of these models 
with a  succession of approximate steps. The approach
follows closely the one used by Biroli and Monasson~\cite{Bimo} for the analysis 
of the spectrum of random matrices. In short, we first define the 
models of interest, we next explain the functional method used
and we finally give a hint on how the equations derived can be solved
in some simple cases.

\subsection{The models}

A family of disordered spin models is defined by the Hamiltonian 
\begin{equation}
H_J = -  
\sum_{i_1 <\dots < i_p} J_{i_1 \dots i_p} s_{i_1} s_{i_2} \dots s_{i_p} 
\; .
\label{model}
\end{equation}
The spins $s_i$, $i=1,\dots, N$, can be Ising variables, $s_i=\pm 1$,
$\forall i$ (Ising model), or they can be real variables. In the latter case one can either constrain them such that $\sum_{i=1}^N s^2_i = N$ (spherical model), or add a ``soft-spin'' term to the Hamiltonian which favors the values of $s_i$ around $\pm 1$. The parameter $p$ is a fixed integer, $p\geq 2$, which controls the number of spins involved in each interaction term. 

The couplings between the spins are given by the quenched 
random variables $J_{i_1 \dots i_p}$.
In the fully-connected  case, all the ${N \choose p}$
entries in these ``tensors'' are 
non-zero and, equivalently, the model 
is defined on the complete (hyper) graph. Generally, one uses a Gaussian 
(or bimodal) probability distribution with zero mean and variance 
$[J_{i_1 \dots i_p}^2]_J=
p! J_0^2/(2N^{p-1})$ in order to ensure a good thermodynamic 
limit. (Henceforth, the square brackets denote an average over quenched 
disorder.) 

In the dilute case, there is only an extensive (proportional to $N$) number of non-zero interaction terms in the sum (\ref{model}).
The interactions are drawn from 
\begin{equation}
P(J_{i_1 \dots i_p}) = \left(1-\frac{\alpha p!}{N^{p-1}}\right) 
\delta(J_{i_1\dots i_p}) + \frac{\alpha p!}{N^{p-1}}
\Pi(J_{i_1 \dots i_p})
\; ,
\label{dist}
\end{equation}
in which $\Pi(J)$ does not contain a Dirac distribution at $J=0$.
The (finite) parameter $\alpha$ is the average ratio of the number of interacting $p$-uplet of spins per variable. 
When $p=2$ this is the Viana-Bray model~\cite{VB,otherVB}. Its geometrical structure is the one of the celebrated  Erd\"os-Renyi random graph~\cite{bollobas}: the model consists of spins occupying the vertices of a graph drawn from this ensemble, with interactions on the edges. For $p \ge 3$ the model is defined on a random hypergraph, the edges are replaced by plaquettes linking $p$ vertices. For any $p$, the probability distribution of the degree of a given spin, i.e. the number of interactions it belongs to, is a Poisson law with mean $\alpha p$.

As there is an extensive number of non-zero terms in (\ref{model}) in the dilute case, each of them must be of order $1$ to obtain a sensible thermodynamic limit. The distribution $\Pi$ of the non-vanishing couplings must thus have finite mean and variance. For concreteness we choose 
\begin{equation}
\Pi(J_{i_1 \dots i_p})=\frac12 \left[ \delta(J_{i_1\dots i_p}-\tilde J)  +
\delta(J_{i_1\dots i_p}+ \tilde J)
\right]
\; ,
\label{dist2}
\end{equation}
with $\tilde J$ finite. 
Note that the fully-connected limit is recovered from the dilute case by taking $\alpha=N^{p-1}/p!$ and ${\tilde J}^2=p! J_0^2/(2N^{p-1})$, or more generally $\alpha \to \infty$, ${\tilde J} \to 0$ with $J_0^2 = 2 \alpha {\tilde J}^2$ finite.

The soft-spin Ising (a) or spherical (b) constraint
are imposed via an additional term in the Hamiltonian:
\begin{equation}
V_I = \sum_{j=1}^N \kappa \, (s^2_j-1)^2
\;\;\; (a) \; ,
\;\;\;\;\;\;\;\;\;\;\;\;
V_S = \mu(t) \sum_{j=1}^N (s^2_j-1) 
\;\;\; (b) \; ,
\label{constraint}
\end{equation}
The (time-independent) 
parameter $\kappa$ should be taken to infinity to recover the Ising limit 
and the (time-dependent) 
Lagrange multiplier $\mu(t)$ is to be determined self-consistently
by imposing that the equal-time correlator be normalised to one.
In what follows we call $H$ the full Hamiltonian, given by the 
sum of $H_J$ and the appropriate constraint, $V=V_I$ or $V=V_S$, 
\begin{equation}
H=H_J + V \; , \qquad  V \equiv \sum_j v(s_j)
\; .
\end{equation}

Our aim is to study the Langevin dynamics of such a model, defined by equations (\ref{dyn}) and (\ref{defnoise}), which mimics the physical situation of a system in contact with a thermostat at fixed temperature.

\subsection{Summary of results}

Several special cases of model (\ref{model})-(\ref{dyn}) have been 
studied in the past. 
A vast majority of these studies has been confined to the (technically simpler) case of fully-connected models. 
For a summary of results see~\cite{yo}. 

There has been however a recent growing interest in dilute 
systems~\cite{VB,otherVB,ModeDo,Mo}, for at least two reasons. On the one hand, even if they are still mean-field, with no notion of geometry, their finite connectivity is an ingredient for the description of real physical systems that was absent from the fully-connected models. Moreover, the fluctuation of their local connectivity leads to Griffiths phases~\cite{Bray-Griffiths}  which are experimentally observed in some physical systems. On the other hand, a large number of optimization problems, satisfiability for instance, can be mapped onto dilute models of Ising spins~\cite{ksat,xor}. This observation triggered a large effort towards the understanding of the static properties of such models, which lead to very interesting recent results~\cite{cavity}. At the moment the understanding of their dynamic behavior is much poorer. One could however hope that it would be very useful for the study and the improvement of local search algorithms that solve these optimization problems~\cite{dynamics-sat}.

Previous studies of physical dynamics of dilute $p$ spin models include:

\noindent
$(i)$ Montecarlo dynamic simulations were performed for 
Ising spins and two-body~\cite{Riccardo,RicciUnpublished} 
or three-body~\cite{Ricci} interactions. 
These numerical studies pointed out the heterogeneous 
character of the non-equilibrium dynamics for fixed quenched disorder.
Moreover Refs.~\cite{Ricci,RicciUnpublished} demonstrated the validity of 
out-of-equilibrium fluctuation-dissipation relations for 
single-spin two-time functions. The single-spin {\sc ofdr}'s turned out to agree with
the results of a static calculation. (See~\cite{ages} for local {\sc fd} relations between 
coarse-grained two-time quantities in finite dimensional glassy systems.)

\noindent
$(ii)$ An analytic solution to the disordered averaged dynamics of 
the spherical $p=2$ dilute model  
can be achieved by solving the Langevin equation in the 
rotated basis in which the interaction matrix is diagonal~\cite{Secu}. This study showed the existence of 
two non-equilibrium dynamic asymptotic regimes. The 
first one is very similar to the one of the fully-connected
counterpart model~\cite{Cude}. The second one is dominated by 
the tails in the spectrum of the random interaction matrix and 
corresponds to progressive condensation on the eigenvectors 
that are localized on these sites. 

Let us state as a side remark here that the spherical dilute model with 
$p\geq 3$ is pathological. Indeed the spherical constraint $\sum_j s_j^2 =N$ can be satisfied in at least two opposite ways. Either all the spins are of order $1$, that is the situation one would like to obtain, or a finite number of them are of order $\sqrt{N}$, a strongly localized situation. Imagine that the $p$ spins interacting through a given plaquette are all localized. As the coupling on a plaquette is of order $1$ in the dilute case, this would contribute with a term of order $N^{p/2}$ to the energy, this situation would thus dominate the thermodynamic limit of such models when $p \geq 3$. Note however that such a pathology can be cured by adding infinitesimal terms of the soft-spin type, $\epsilon \sum_j (s_j^2 - 1)^n$, with $n$ sufficiently high,  to the Hamiltonian.

\subsection{Functional formalism}

In this Section we introduce a generic formalism to 
derive macroscopic dynamic equations for the 
evolution of global correlation and response functions.

Standard techniques allow us to derive 
a dynamic generating functional, ${\cal Z}$, for a generic Langevin 
process with white noise, as a path integral~\cite{MSR}:

\begin{eqnarray}
{\cal Z}[\eta,\hat\eta] &=& 
\int {\cal D}s_i {\cal D}i\hat s_i 
{\cal D} \overline \psi_i {\cal D} \psi_i \; \exp(-S_{\sc eff} + \int_0^t dt' \; 
\left[ \eta_i(t') s_i(t') -  \hat \eta_i(t') i \hat s_i(t')\right]) 
\nonumber\\
-S_{\sc eff} &=& 
 \int_0^t dt' \left[
T (i\hat s_i(t'))^2 
+ i\hat s_i(t') \left(\partial_t s_i(t') + \frac{\delta H}{\delta s_i(t')}
\right)
\right]
\nonumber\\
& & 
+
\int_0^t dt' \int_0^t dt'' \;
\overline\psi_i(t') 
\left( \delta(t'-t'') \delta_{ij} \partial_{t''} + 
\frac{\delta^2 H}{\delta s_i(t') s_j(t'')}\right)  \psi_j(t'')
\label{Seff}
\end{eqnarray}
where we have introduced two time-dependent sources $\eta_i(t)$ and 
$\hat\eta_i(t)$. Einstein's summation convention is assumed. The fields $s_i$ and $\hat{s}_i$ are real, $\hat{s}_i$ being a response field conjugated to $s_i$, whereas $\overline\psi_i$ and $\psi_i$ are Grassman fermionic fields. The correlation and response functions of the stochastic process solution of the Langevin equation (\ref{dyn}), averaged over the different realizations of the noise (\ref{defnoise}), are given by derivatives of ${\cal Z}$:
\begin{eqnarray}
&&
C(i_1,t_1,\dots,i_k,t_k) = \frac{\delta^k {\cal Z}}{\delta \eta_{i_1}(t_1) \dots \delta \eta_{i_k}(t_k)} \;, \\
&&
R(i_1,t_1,\dots,i_{k-1},t_{k-1}; i_{k},t_k) = \frac{\delta^k {\cal Z}}{\delta \eta_{i_1}(t_1) \dots \delta \eta_{i_{k-1}}(t_{k-1}) \delta \hat{\eta}_{i_k}(t_k)} \;, \\
&& R(i_1,t_1,\dots,i_{k-2},t_{k-2}; i_{k-1},t_{k-1},i_{k},t_k) =\nonumber\\ 
&&\;\;\;\;\;\;\;\;
\;\;\;\;\;\;\;\;\;\;\;\;\;\;\;\;
\frac{\delta^k {\cal Z}}{\delta \eta_{i_1}(t_1) \dots \delta \eta_{i_{k-2}}(t_{k-2}) \delta \hat{\eta}_{i_{k-1}}(t_{k-1}) \delta \hat{\eta}_{i_k}(t_k)} \;,
\end{eqnarray}
the sources being set to zero after calculating the derivatives. In other words, the correlation and response functions are averages taken with the normalized weight $\exp(-S_{\sc eff})$. The response fields $- i \hat{s}_i(t)$ are inserted at the times where a magnetic field $h_i$ has perturbed the system.

${\cal Z}$ can be written in a very compact form 
if one introduces the super-field formulation of stochastic processes
as explained in \cite{susy}. In this approach
one first enlarges (space)-time to include two Grassmann coordinates
$\theta$ and $\overline\theta$, {\it i.e.}
$t\to a=(t,\theta,\overline\theta)$.
The dynamic variables $s_i(t)$ and the auxiliary variable 
$i \hat s_i(t)$ together with the fermionic ones $\psi_i(t)$ and
$\overline\psi_i(t)$ are encoded  in a super-field, 
\begin{equation}
\Phi_i(a) = 
s_i(t) +  \overline\theta \psi_i(t) + \overline\psi_i(t) \theta
+ i \hat s_i(t) \theta\overline\theta 
\; .
\label{susy-field}
\end{equation}
With these definitions, 
\begin{eqnarray}
&&{\cal Z}[\zeta] = \int 
{\cal D}\vec\Phi \exp\left(\frac12 \int da \; \Phi_i(a) 
D_{a}^{(2)} \Phi_i(a)-\int da \; H[\vec\Phi(a)] + \int da \; \Phi_i(a) 
\zeta_i(a) \right)
\nonumber\\
&&
\label{generating-susy}
\end{eqnarray} 
with 
$da\equiv dt \, d\theta \, d\overline\theta$, 
$\vec\Phi \equiv (\Phi_1,\dots,\Phi_N)$, 
$\zeta_i(a) \equiv \eta_i(t) \overline\theta\theta + \hat \eta_i(t)$,  
and the dynamic operator $D_a^{(2)}$ is defined as 
\begin{equation}
-D_a^{(2)}= 
2 T \frac{\partial^2}{\partial \theta\partial\overline\theta}
+ 2 \theta \frac{\partial^2}{\partial \theta\partial t} 
- \frac{\partial}{\partial t} 
\; .
\end{equation}

If the model is spherically constrained, one can 
absorb the constraint in the quadratic part of the 
action by redefining
\begin{equation}
D_a^{(2)} \to D_a^{(2)} + \mu_s(a)
\end{equation} 
with $\mu_s(a)$ a super Lagrange multiplier, with the form
(\ref{susy-field}).
If one deals, instead, with the soft spin model, the non-quadratic
term $V_I[\Phi]$ appears in $H$.

The super-symmetric  notation allows one to encode in the 
single super correlator
\begin{equation}
Q_{ij}(a,b) \equiv \langle \Phi_i(a) \Phi_j(b)\rangle
\end{equation}
the two-time correlators and responses, $C_{ij}(t, t')$ and $R_{ij}(t, t')$. Indeed, the super-correlator $Q_{ij}(a,b)$  has 16 ``components''. However, 
many of these vanish or are related to each other due to 
the symmetries that should hold in a physical 
situation~\cite{susy} as ghost number
conservation and causality. Imposing these properties one is 
left with a simpler expression for $Q_{ij}$ that encodes the physical self-correlation and the linear response:
\begin{equation}
Q_{ij}(a,b) = C_{ij}(t_a,t_b) + (\overline \theta_b - \overline \theta_a) 
\left( \theta_b R_{ij}(t_a,t_b) - \theta_a R_{ji}(t_b,t_a)\right)
\; .
\label{susyQ}
\end{equation}

Note the terms in which two fermions appear equal 
the response functions. 
The equilibrium properties can also be expressed as super-symmetries and constrain further the two-point super-correlator imposing {\sc tti} and {\sc fdt}. 

Similarly, all the three point functions defined in Section~\ref{sectionmanypoint} are contained in
$Q_{ijk}(a,b,c) \equiv \langle \Phi_i(a) \Phi_j(b) \Phi_k(c)\rangle$.

We are interested in correlations and responses averaged over the disorder. The fact that $\cal Z$ is equal to $1$ in the absence of sources, independently of the realization of the disorder~\cite{cirano}, implies that these averaged functions are generated by $[{\cal Z}]_J$. Since the generating functional itself 
(and not its logarithm) has to be averaged, one can  avoid the 
use of replicas. The only term that depends on disorder 
in  the action in $\cal Z$ is $H_J$. The average over 
the probability distribution of the couplings (\ref{dist}) and (\ref{dist2})  reads
\begin{eqnarray}
&&
\left[ e^{ -\int da H_J[\vec\Phi(a)] } \right]_J \equiv e^{-N H_{\sc eff}} 
\nonumber\\
&&
\;\;\;\;\;\;
= \prod_{i_1<\dots<i_p} \left[ 1 + \frac{\alpha p!}{N^{p-1}} \left( \cosh \left( {\tilde J} \int da \Phi_{i_1}(a) \dots \Phi_{i_p}(a) \right) - 1 \right) \right] 
\nonumber\\
&& 
\;\;\;\;\;\;
=
\exp \left[ N \alpha \frac{1}{N^p} \sum_{i_1 \dots i_p} \left( \cosh \left( {\tilde J} \int da \Phi_{i_1}(a) \dots \Phi_{i_p}(a) \right) - 1 \right) \right]\label{heff_dilue}
\end{eqnarray}
at leading order in $N$. 

The next step to take is to disentangle this effective Hamiltonian
and  to reduce the full problem to a single-spin one, that will be evaluated using saddle-point methods. An advantage of the {\sc susy} notation is that it parallels the static replica calculations~\cite{jorge}.
 The manipulation of the dynamic effective Hamiltonian  is done 
following the same steps as  in the replica static calculation. For this one takes advantage of the empiric correspondence between super-coordinates and replica indices. 

For fully-connected models, the reduction to a single spin problem can be done  by introducing a {\it global}  two point function $Q(a,b)$, or the {\it global} Parisi matrix $Q_{ab}$ in replica terms. For dilute models the procedure is much more difficult. One possibility is to introduce the whole set of multi-point super-correlators $Q(a,b)$, $Q(a,b,c)$, etc. This route was followed in the original treatment of the statics of the Viana-Bray model~\cite{VB}. We follow here a 
different strategy that  was introduced by Monasson~\cite{Mo} (see also \cite{ModeDo}) in the replica static context. It is based upon the introduction of a functional order parameter.

Let us define $c(\Psi)$ as the fraction of sites with super-field
$\Phi_i$ identical to a chosen value $\Psi$
\begin{equation}
c(\Psi) \equiv \frac1N \sum_{i=1}^N  \delta \left[ \Psi -\Phi_i \right]
\; ,
\label{cPhi-def}
\end{equation}
where the functional $\delta$ enforces that $\Psi(a) = \Phi_i(a)$ for all values of the super-coordinate $a$. 
Note that $c$ is normalised, 
$\int {\cal D}\Psi \, c(\Psi) =1$.

As emphasized in~\cite{Mo}, the mean-field character of the models implies that their effective Hamiltonian, after performing the average over disorder, can be expressed in terms of such a {\it global} functional order parameter. For the dilute $p$ spin models under consideration here, one obtains
\begin{equation}
-H_{\sc eff}[c] = \alpha \int {\cal D}\Psi_1 \dots {\cal D}\Psi_p \; c(\Psi_1) \dots c(\Psi_p)
\left[ \cosh \left( {\tilde J} \int da \; \Psi_1(a) \dots \Phi_p(a) \right) - 1 \right]
\; .
\nonumber\\
\label{heff}
\end{equation}

We enforce  the
 definition of $c(\Psi)$ in the generating functional by 
introducing an identity in its path integral representation: 
\begin{eqnarray}
1 &=& 
\int {\cal D} c {\cal D} i\hat c \;  
\exp \left[\int {\cal D}\Psi \, i \hat c(\Psi) 
( N c(\Psi) - \sum_{i=1}^N 
\delta\left[\Psi - \Phi_i\right] ) \right]
\nonumber\\
&=&
\int {\cal D}c \, {\cal D}i\hat c \; 
\exp\left[\int {\cal D}\Psi N i \hat c(\Psi) c(\Psi) -
\sum_{i=1}^N i \hat c(\Phi_i) \right]
\; .
\end{eqnarray}
After doing so we achieved our goal of disentangling the $N$ degrees of freedom 
\begin{equation}
[{\cal Z}]_J = 
\int {\cal D}c {\cal D}i \hat c  \; \exp\left[-N G\right] 
\end{equation}
with 
\begin{equation}
G =
- \int {\cal D}\Psi \; i \hat c(\Psi) c(\Psi)
+ H_{\sc eff} 
- \ln \left[ \int {\cal D}\Psi \; e^{\frac{1}{2} \int da \Psi(a) D_a^{(2)} \Psi(a) - \int da v(\Psi(a)) - i\hat{c}(\Psi)} \right]
\; .
\end{equation} 

We can now evaluate $\cal Z$ with the saddle-point method.
Enforcing that $G$ is stationary with respect to $c$ and $\hat{c}$  yields:
\begin{eqnarray}
c^{\sc sp}(\Psi) &=& \lambda \; \exp \left[ \frac{1}{2} \int da \; \Psi(a) D_a^{(2)} \Psi(a) - \int da \; v(\Psi(a)) - i\hat{c}^{\sc sp}(\Psi)\right] \; , \label{saddle1}\\ 
\hat{c}^{\sc sp}(\Psi) &=& \left. \frac{\delta H_{\sc eff}[c]}{\delta c(\Psi)} \right|_{c^{\sc sp}} \; ,\label{saddle2}
\end{eqnarray}
with $\lambda$ a normalization constant.

Let us give the interpretation of such equations. The correlation functions computed with the normalized weight $c^{\sc sp}(\Psi)$ are the averages of the original single-site functions:
\begin{equation}
\int{\cal D} \Psi \; c^{\sc sp}(\Psi) \;  \Psi(a_1) \dots \Psi(a_k) = \frac{1}{N} \sum_i \left[ \langle \Phi_i(a_1) \dots \Phi_i(a_k) \rangle \right]_J \; .
\end{equation}
To enlighten the physical meaning of Eqs.~(\ref{saddle1}) and (\ref{saddle2}), let us consider the fully connected limit, with $\alpha \to \infty$, ${\tilde J} \to 0$ and $J_0^2 = 2 \alpha {\tilde J}^2$ finite.
Expanding the hyperbolic cosine in (\ref{heff}), only the first term of the series survives:
\begin{equation}
H_{\sc eff} \to \frac{J_0^2}{4} \int {\cal D}\Psi_1 \dots {\cal D}\Psi_p \; c(\Psi_1) \dots c(\Psi_p) \int da db \; \Psi_1(a) \Psi_1(b) \dots \Psi_p(a) \Psi_p(b)
\label{heff_limit}
\end{equation}
The relation between $\hat{c}^{\sc sp}$ and $c^{\sc sp}$ thus becomes:
\begin{equation}
\hat{c}^{\sc sp}(\Psi) = \frac{p J_0^2}{4} \int da db \; Q^{\sc sp}(a,b)^{\bullet (p-1)} \Psi(a) \Psi(b) \; , \quad Q^{\sc sp}(a,b) \equiv \int {\cal D}\Psi \; c^{\sc sp}(\Psi) \;  \Psi(a) \Psi(b)
\end{equation}
and $\bullet$ represents the direct or Hadamard product
(see the Appendix~\ref{convolution}). We thus obtain the {\sc susy} form of the well-known single spin equation for fully connected models: averages of $\Psi$ are taken with a Gaussian weight (apart from the soft-spin  $v$ term), to be determined self-consistently through $Q^{\sc sp}$. It is the equation of a single degree of freedom evolving through a Langevin equation with a retarded interaction and a colored noise, which are to be expressed self-consistently in terms of the correlation and the response of the process~\cite{Sozi}.

In the dilute case, one can also follow this idea, but the self consistency equations cannot be written solely in terms of the two point function. The whole hierarchy of many-coordinate correlations appears:
\begin{eqnarray}
&&c^{\sc sp}(\Psi)=\lambda \, \exp \left[ \frac{1}{2} \int da \; \Psi(a) D_a^{(2)} \Psi(a) - \int da \; v(\Psi(a)) \right. 
\;\;\;\;\;\;\;\;\;\;\;\;\;\;\;\;\;\;\;\;\;\;\;\;
\nonumber \\
&& 
\;\;\;\;\;\;\;\;\;\;\;\;\;\;\;
\left. -\alpha p \sum_{n=1}^{\infty} \frac{\tilde{J}^{2 n}}{(2n)!} \int da_1 \dots da_{2n} \; \Psi(a_1) \dots \Psi(a_{2n}) \;  Q_{2n}(a_1,\dots,a_{2n})^{\bullet (p-1)} \right] \; ,
\label{series1}\\
&&
Q_{2n}(a_1,\dots,a_{2n}) \equiv \int {\cal D}\Psi \; c^{\sc sp}(\Psi) \;  \Psi(a_1) \dots \Psi(a_{2n})
\label{series2}
\end{eqnarray}

This equation cannot be solved exactly. A set of approximation
schemes can be envisaged. In the following we critically 
discuss some of the approximations one can use. 

\subsection{Approximation scheme}

In this Subsection we discuss some approximations to the 
saddle-point equations (\ref{series1}) and (\ref{series2}) that is exact in the thermodynamic limit and determines the behavior of all 
global correlators. 

\subsubsection{Cutting the series}

The simplest approximation one can envisage is to 
simply cut the series appearing 
in the right-hand-side of (\ref{series1}) after the first term
in such a way that only $Q(a,b)$ enters the approximated
equation. This is equivalent to  proposing that $\hat{c}$ is quadratic. 
It is easy to see that with this approximation we recover the fully-connected counterpart model, 
with $J_0=\tilde{J} \sqrt{2 \alpha}$. The solution does not go
beyond the results already known for this case~\cite{Cuku1,Cuku2,Cude}.

Presumably, better approximations are obtained by progressively keeping some higher order correlations. More precisely, 
one could cut the series after the second term and then 
derive (exactly in the spherical case or 
with a further approximation to treat the effect of the soft-spin term $v$)  a set of dynamic 
equations coupling $Q(a,b)$ and $Q(a,b,c,d)$ only.
One could also cut the series after the third term and then derive a 
set of dynamic equations coupling $Q(a,b)$, $Q(a,b,c,d)$
and  $Q(a,b,c, d,e,f)$ and so on and so forth.

This is the kind of 
approach used when deriving and cutting {\sc bbgky} 
hierarchies in field-theoretical models or condensed-matter 
problems. It is also similar to the procedure used by Viana and 
Bray~\cite{VB} in their study of the statics of the dilute 
$\pm$ spin-glass 
model with two-body interactions close to its transition temperature where the $Q$'s (in replica space) can be assumed to be small.

\subsubsection{An iterative procedure}

A different procedure consists in evaluating $c(\Psi)$  (and the super correlators) iteratively. The idea is:

\noindent
({\it i}) In the first step one uses the simplest {\it Ansatz} for 
$c(\Psi)$ and evaluates $H_{\sc eff}$. 
(We discuss two ways of implementing this initial step).

\noindent
({\it ii})  Next, one uses Eq.~(\ref{saddle1}) [after replacing $\hat c$ with (\ref{saddle2})] to obtain an improved 
expression for $c^{\sc sp}(\Psi)$. 

\noindent
({\it iii}) With this form one calculates $H_{\sc eff}$ again and goes 
to ({\it ii}).

This procedure follows very closely the proposal of 
Biroli and Monasson to compute the density of states of 
sparse random matrices~\cite{Bimo}. The {\sc  susy}
formulation of the dynamic generating functional makes the
treatment of the dynamic problem very similar to the static
calculation that uses the replica trick (see~\cite{jorge}
for a recent discussion of the relation between {\sc susy} and replica
analysis). 

An additional feature to be considered in this iteration is how we 
update the constraint on the normalization of the correlation,
{\it i.e.} the fact that $Q(a,a)=1$. For the spherical model we shall 
demand this constraint to be valid at each step of the iteration.
This means that we shall modify the Lagrange multiplier in such 
a way to impose the constraint.

Importantly enough, after the first iteration one accesses
the functional order parameter $c(\Psi)$ and hence  
{\it all} many-point correlations simultaneously though approximately. One can expect that even after using only
one step of the iteration this method will yield better information 
than simply cutting the series keeping only a finite number of 
terms. This is indeed the case when studying the spectral properties of random matrices. 

\subsubsection{First step: the effective medium approximation}

In the first step we use the simplest approximation 
that captures the same behavior as the ``effective medium
approximation''~\cite{Bimo,Secu2}.
In this approximation, one treats the environment of each point 
on the random graph in a uniform manner. In the calculation of the 
spectral density of sparse random matrices this approximation yields a
symmetric distribution with a finite support that qualitatively 
resembles a semi-circle. Thus, the result is a simple modification 
with respect to the usual Gaussian 
case. This approximation fails to capture the effect of highly 
connected sites that clearly deviate from  the effective medium 
assumption. These imply the appearance of tails in the density
of states that are not obtained at this level of the calculation.

The dynamic effective medium approximation  can be done in at least two ways:

\vspace{0.2cm}
\noindent{\it Cutting the series}
\vspace{0.2cm}

Going back to what we discussed above, one can use as the 
starting $c(\Psi)$ the result of cutting the series in (\ref{series1}) 
after the first term, deriving and solving  a
self-consistent equation for $Q(a,b)$ and  using this as an input 
to compute $c(\Psi)$. 

\vspace{0.2cm}
\noindent{\it Gaussian approximation}
\vspace{0.2cm}

A slightly different (but qualitatively equivalent) starting point 
is given by  proposing a Gaussian {\it Ansatz}  for $c(\Psi)$:
\begin{eqnarray}
c_{\sc ema}(\Psi) &=& 
({\det Q})^{-1/2}
\; \exp\left(-\frac12 \int da db \; \Psi(a) Q^{-1}(a,b) \Psi(b)\right)
\label{Gaussian-ansatz} 
\; .
\label{gauss-ansatz}
\end{eqnarray}
One can easily check that 
the denominator
ensures the normalization of  $c(\Psi)$ and 
that $Q(a,b)$ is correctly given by 
\begin{eqnarray*}
Q(a,b) = \frac1{\sqrt{\det Q}}
\int {\cal D}\Psi \, \Psi(a) \Psi(b)
\, \exp\left(-\frac12 \int da' db'
\; \Psi(a') Q^{-1}(a',b') \Psi(b') \right)
\; .
\end{eqnarray*}
This form is closer to the approximation used in~\cite{Bimo,Secu2}.

To advance further it is convenient to express the 
$c$-dependent effective action in terms of $Q(a,b)$.
This can be done exactly for the spherical problem. When we
deal with the soft spin case, instead, the non-quadratic 
term $V_I(\Psi)$ has to be treated within an additional 
approximation. We shall discuss this point later. For the 
moment we focus on the spherical models.
After rather simple manipulations 
one finds 
\begin{eqnarray*}
[{\cal Z}]_J &=& \int {\cal D}Q \; e^{-S_{\sc eff}(Q)} \; , \;\;\;\;\; \mbox{with} 
\nonumber\\ 
2S^{\sc sph}_{\sc eff}(Q) &= &
\mbox{Tr} \ln Q +
\int da db \; \delta(a-b) (D_a^{(2)}+\mu_s(a)) Q(a,b) 
- 2H^J_{\sc eff}(Q)
\; . 
\end{eqnarray*} 
The superscript $J$ in $H_{\sc eff}$ indicates that 
the effective Hamiltonian depends only on the disorder
dependent term $H_J$. 
Its variation with respect to $Q$ yields:
\begin{equation}  
\frac{\delta S^{\sc sph}_{\sc eff}(Q)}{\delta Q} = 0 =
\frac12 Q^{-1}(a,b) + \frac12 (D_a^{(2)}+\mu_s(a)) \delta(a,b) 
- 
\frac{\delta H^J_{\sc eff}(Q)}{\delta Q(a,b)}
\; . 
\end{equation}
Multiplying this equation operationally 
by $Q(b,a')$ (see Appendix~\ref{convolution})
the dynamic equation takes the more
familiar Schwinger-Dyson form
\begin{equation}
(D_a^{(2)}+\mu_s(a)) Q(a,b)+\delta(a-b) + 
\int da' \; \Sigma(a,a') Q(a',b) = 0
\; ,
\label{MCeq2}
\end{equation}
with 
\begin{equation} 
\Sigma(a,a') = -2 \frac{\delta H_{\sc eff}(Q)}{\delta Q(a,a')}
\; .
\label{self-energy-dilute}
\end{equation}
As mentioned above, we require that $Q$ satisfy the spherical 
constraint. Thus, we fix the evolution of $\mu_s(a)$ to 
ensure the validity of this constraint.

It is now instructive to see how one recovers the 
well-known equations for the fully-connected model.
We have seen in Eq.~(\ref{heff_limit}) how the effective Hamiltonian simplifies in this limit. Plugging in the Gaussian {\it Ansatz} (\ref{Gaussian-ansatz}), 
$H^J_{\sc eff}$ becomes a simple
function of $Q(a,b)$,
\begin{equation}
H^J_{\sc eff}(Q) = \frac{J_0^2}4 \int da db \;Q^{\bullet p}(a,b)
\; ,
\label{Heff-fc}
\end{equation}
and the self-energy $\Sigma(a,b)$ is easily deduced from this expression.
A way to prove 
that the Gaussian {\it Ansatz} (\ref{Gaussian-ansatz})
is exact for the fully-connected case is to 
check that the {\it exact} equation for $c(\Psi)$ coincides with the 
one obtained from the Gaussian {\it Ansatz} and the saddle-point 
evaluation.

In the dilute case, one has to distinguish between the cases $p=2$ and $p \geq 3$. Indeed, consider the expression~(\ref{heff}), where one uses the Gaussian {\it Ansatz} (\ref{Gaussian-ansatz}). The integrals over $\Psi$ can be formally performed with an expansion of the hyperbolic cosine and the use of Wick's theorem. When $p=2$, all the terms can be written explicitly and the asymptotic behavior of the two-point correlator can be worked out. We do not reproduce the calculations here as they do not give more insight into the physics of the problem: as expected, at this level of approximation one finds back a behavior typical of the fully-connected counterpart model, as was shown in a more direct way in~\cite{Secu}. For $p\geq 3$, $H_{\sc eff}^J$ can only be expresssed as  a series expansion involving the 
supercorrelators. The properties of the Hadamard and 
operational products explained in the Appendix should allow us to 
prove that the solution to this equation has the same properties as the one for the fully-connected $p=3$ model. 

Even if the {\sc ema} equations found with the Gaussian {\it Ansatz} are much more complicated than the 
one obtained by cutting the series, one can check that 
it contains the same qualitative information, which is the one of the fully connected counterpart model.

\vspace{0.2cm}
\noindent{\it Soft spin models}
\vspace{0.2cm}

The term $v_I(\Psi)=a(\Psi^2-1)^2$ in $H$ is not quadratic
and one has to resort to an additional approximation to treat
it. One possibility is to identify the simplest non-trivial term
that it generates in a series expansion around the critical 
temperature at which one expects the order parameter to vanish,
{\it i.e.} a second order phase transition. This is the proposal 
in \cite{truncated}, later used in the study of the dynamics of the 
Sherrington-Kirkpatrick model in \cite{Cuku2}. Another possibility,
used in \cite{claudio}, 
is to treat this term in the mode-coupling approximation.
The terms obtained with the first and the second procedure are
slightly different. Even if they modify the details of the 
dynamic solution in the fully-connected case, they do not modify
the qualitative features of it. We expect these same generic
features also in the dilute case.

\subsubsection{Second step: the single defect approximation}

In the study of the density of states of
sparse random matrices one can go beyond 
the Gaussian approximation~\cite{Bimo,Secu2}. 
The same kind of approach can be used in the analysis of the 
dynamics of dilute disorder models. The idea is to use 
Eqs.~(\ref{saddle1}) and~(\ref{saddle2}) iteratively. In the first step one 
replaces $c_{\sc ema}(\Psi)$ on the right-hand-side of 
(\ref{saddle2}) and computes its improved functional form.
In the case of matrices this gives access to the tails in the 
eigenvalue distribution. We expect that this improved approximation 
will also modify the result for the dynamic behavior considerably
since it will capture, at least partially, the effect of 
heterogeneities in the connectivity of the sites in the random
(hyper) graph.

In order to clarify the way in which the iteration is implemented,  let us neglect the soft-spin term (this, of course, is not correct if we are treating an Ising spin problem in which case, as already said,  we have to treat 
$v$ with a further approximation).
Thus, $H_{\sc eff}=H_{\sc eff}^J$ and in 
order to compute the first iteration we insert
\begin{equation}
\frac{\delta H^J_{\sc eff}}{\delta c(\Psi_1)}
= -\alpha p  + \alpha p \int {\cal D}\Psi_2 \dots {\cal D}\Psi_p
\; c(\Psi_2) \dots c(\Psi_p) \cosh\left( \tilde J \int da \Psi_1(a) \dots  
\Psi_p(a) \right)
\end{equation}
in the right-hand-side of Eq.~(\ref{saddle1}) and expand the last term as
\begin{eqnarray}
& & c_{\sc sda}(\Psi_1) = 
\lambda' 
e^{\frac12 \, \int da \Psi_1(a) [D_a^{(2)} +\mu_s(a)] \Psi_1(a)}
\sum_{k=0}^\infty \frac{e^{-\alpha p} (\alpha p)^k}{k!} 
\nonumber\\
&&
\times \left[
\int {\cal D}\Psi_2 \dots {\cal D}\Psi_p \; c_{\sc ema}(\Psi_2) \dots 
c_{\sc ema}(\Psi_p) 
\cosh\left( \tilde J \int da \Psi_1(a) \dots \Psi_p(a) \right) \right]^{\bullet k}
\; .
\nonumber\\
\end{eqnarray}
For the sake of clarity we concentrate on the $p=2$ case.
Using now the Gaussian expression for $c_{\sc ema}(\Psi)$ with variance $Q_{\sc ema}$, we find
\begin{eqnarray}
c_{\sc sda}(\Psi) &=&
\lambda'
\sum_{k=0}^\infty \frac{e^{-2 \alpha} (2 \alpha)^k}{k!} \; 
\exp\left[-\frac12 \, \int da db \; \Psi(a) A_k(a,b) \Psi(b)\right]
\end{eqnarray} 
with 
\begin{equation}
A_k(a,b) \equiv - \delta(a-b) [D_a^{(2)} + \mu_s(a) ] - k \tilde J^2
Q_{\sc ema}(a,b)
\label{Adef}
\end{equation}
Rewriting this equation in the form
\begin{equation}
\delta(a,b) \equiv [D_a^{(2)} + \mu_s(a) ] A_k^{-1}(a,b)- k \tilde J^2
\int db' \, Q_{\sc ema}(a,b')A_k^{-1}(b',b)
\label{Qdef2}
\end{equation}
we realize that this is a Schwinger-Dyson equation for a ``system''
(the unknown $A_k^{-1}$) in the presence of a complex external bath:
a white bath giving rise to the dynamic operator $D_a^{(2)}$ and a 
``colored bath'' function of the {\sc susy} self-energy 
$Q_{\sc ema}$~\cite{japan,yo}.
The latter has a slow relaxation, of glassy type, since
in the {\sc ema} the dilute model behaves in a very similar manner
to its fully-connected relative.
This means that the {\sc susy} correlator
$Q_{\sc ema}$ encodes a correlation and a response with two two-time
regimes, a stationary and an aging one, the former controlled by the 
temperature of the external bath and the latter characterized by 
an effective temperature that differs from the one of the environment.
If we assume --- a hypothesis that has to checked! --- that the effect 
of this colored bath on the system is as discussed in \cite{japan,yo},
the latter will follow the dynamics dictated by the complex bath
with two time-scales and two values of the effective temperature. 

The spirit of the iteration is to use the previously determined 
value of $Q(a,b)$ as an input. For the spherical model 
$\mu_s(t)$ is also a function that is self-consistently determined 
to impose the spherical constraint at equal times. 
We believe that a convenient way to deal with the constraint is to impose it on each step of the approximation. Thus, we let the 
value of $\mu_s(t)$ appearing in the {\sc sda} level of the calculation free and we fix it by imposing that the correlations 
at equal times be normalized to one.
To this end,  it is convenient to impose the spherical condition
on $A_k^{-1}$ since this ensures, automatically, the normalization 
of $Q_{\sc sda}$ at equal times with $\lambda'=1$. 

Now, using $Q_{\sc sda}(a,b) \equiv \int {\cal D} \Psi \; \Psi(a) \Psi(b) c_{\sc sda}(\Psi)$,  we have 
\begin{equation}
Q_{\sc sda}(a,b) = 
\lambda'
\sum_{k=0}^\infty \frac{e^{-2 \alpha} (2 \alpha)^k}{k!} \;
\frac{A_k^{-1}(a,b)}{\sqrt{\det A_k}}
\; .
\label{sda-eq}
\end{equation}
One recognizes in (\ref{sda-eq}) 
a series with the same structure  as Eq.~(29) in 
\cite{Secu2}. As explained in this reference, this series captures the 
effect of the heterogeneities in the connectivity of sites in the 
random graph. Each term represents the contribution of sites connected 
to $k$ neighbors. The {\sc sda} super correlator is then a 
very different entity from the {\sc ema} one. It is 
given by the ``superposition'' of independent aging systems 
(labeled by $k$) each weighted with the Poissonian distribution.

The complete solution of the problem, at this step, 
needs more work. We should be able to estimate the 
behavior of this two-time dependent series but this is 
not an easy task. It would be very interesting to check
how the results in \cite{Secu} are recovered with this approach. 
In particular, a property of this solution that is not easily
seen from  the calculations above  is the fact that the asymptotic
value of the Lagrange multiplier actually diverges when $p=2$.

\section{Conclusions}

In this paper we discussed several aspects of the dynamics of 
disordered spin models. 

The first part is general and deals 
with the generic properties of many-point functions
in equilibrium and the modifications expected out of equilibrium 
for models with slow dynamics. In particular we formulate
an explicit conjecture, {\it cfr.} Eq.~(\ref{GeneralizedFluctuationPrinciple}), 
for the out-of-equilibrium
{\sc fd} relations relating the multi-time correlation and response 
functions. This conjecture proves to be correct for gaussian
aging processes, and should be amenable for a numerical check.

In the second part of the paper 
we presented an approach to the disordered
averaged dynamics of dilute random spin models. Since with this 
approach we study the dynamic generating functional
averaged over disorder we  only have access to typical 
properties and we cannot follow the single spin dynamics in the 
particular background of a quench disorder realization.
This should be contrasted with the recent advances in static calculations
which allows a detailed sample-by-sample analysis of similar models~\cite{cavity,Ricci}. 

When making contact between the generic method 
we discussed here and the explicit solution to the 
spherical dilute spin-glass model with two-body interactions~\cite{Secu}
we see that with each level of the approximation we access
different time-scales. Indeed, with the first {\sc ema} we only
see the dynamics in a first long-times scale that resembles 
strongly the fully-connected  partner of the model. With this 
calculation the fluctuations in the connectivities are averaged 
out and the result is consistent with it. One step beyond one 
sees the fluctuations in the number of first neighbors of 
the sites and the disordered-average dynamics feels the 
existence of the special sites with larger connectivity than the 
average one, in a longer (still aging) time-scale. 
It would be interesting to understand how the existence and number of these 
time-scales and/or the 
behavior of the system in them is modified  when including more
levels in the iteration.

\appendix

\section{Properties of super correlators}
\label{convolution}

Two products between {\sc susy} correlators can be defined. These are 
the convolution and the Hadamard product:
\begin{eqnarray}
Q^{(1)}(a,b) \otimes Q^{(2)}(b,c) &\equiv& \int db \; Q^{(1)}(a,b) Q^{(2)}(b,c) 
\label{conv-prod}
\;,
\nonumber\\
Q^{(1)}(a,b) \bullet Q^{(2)}(a,b) &\equiv & Q^{(1)}(a,b) Q^{(2)}(a,b) 
\;.
\label{Hadamard}
\end{eqnarray}
These products are associative and commutative.

If one multiplies two {\sc susy} correlators of the form
(\ref{susyQ}) with either of these products one obtains
another {\sc susy} correlator of the same form. Its new ``components'' 
are given by functionals of the components of the original 
susy correlators. Indeed, 
\begin{eqnarray}
Q(a,b) = (Q^{(1)} \otimes Q^{(2)}) (a,b)  =
C(t_a,t_b) + (\overline \theta_b - \overline \theta_a) 
\left( \theta_b R(t_a,t_b) - \theta_a R(t_2,t_1)\right)
\end{eqnarray}
with 
\begin{eqnarray}
C(t_a,t_b) &=& \int dt_c \; 
[ R^{(1)}(t_a,t_c) C^{(2)}(t_c,t_b) +
C^{(1)}(t_a,t_c) R^{(2)}(t_b,t_c) ]
\; ,
\nonumber\\
R(t_a,t_b) &=& \int dt_c \; 
R^{(1)}(t_a,t_c) R^{(2)}(t_c,t_b)
\; ,
\end{eqnarray}
and 
\begin{eqnarray}
Q(a,b) = (Q^{(1)} \bullet Q^{(2)}) (a,b)  =
C(t_a,t_b) + (\overline \theta_b - \overline \theta_a) 
\left( \theta_b R(t_a,t_b) - \theta_a R(t_2,t_1)\right)
\end{eqnarray}
with 
\begin{eqnarray}
C(t_a,t_b) &=& C^{(1)}(t_a,t_b) C^{(2)}(t_a,t_b)
\; ,
\nonumber\\
R(t_a,t_b) &=&  
R^{(1)}(t_a,t_b) C^{(2)}(t_a,t_b) +  
C^{(1)}(t_a,t_b) R^{(2)}(t_a,t_b) 
\; .
\end{eqnarray}

It is very simple to check that time-translation invariance
and the fluctuation-dissipation relation are conserved by 
these products. More precisely, if one multiplies two
super correlators satisfying these properties, the results will 
also verify them. Moreover, multiplying two super-correlators
evaluated in times that are sufficiently far away such that they 
evolve in their slow (aging) scale, the result will also be slow.
This property is apparent for the Hadamard product. To prove it
for the convolution one needs to separate the integration 
time in several intervals and approximate the integrals as 
usually done when solving the dynamics of fully-connected models
(see {\it e.g.}~\cite{yo}). By induction one can then prove that 
any $k$-th power in the Hadamard or convolution sense
of a {\sc susy} correlator in the 
fast, stationary regime yields a result within this regime. 
Similarly, any Hadamard or convolution power of {\sc susy} correlators
in the slow, aging regime stays in the same regime. This 
property is very useful for solving the {\sc ema} equations.

\vspace{0.5cm}
\noindent{\underline{Acknowledgments}}
L. F. C. is research associate at ICTP Trieste and a Fellow of the 
Guggenheim Foundation. We acknowledge financial 
support from the ACI Jeunes Chercheurs ``Algorithmes 
d'optimisation et syst\`emes d\'esordonn\'es quantiques''.
This research was supported in part by the National Science Foundation under Grant No. PHY99-07949.

\end{document}